\documentclass[twocolumn]{aastex631}

\defcitealias{park21}{P21}

\begin{document}

\title{Mid-Infrared Variability of Young Stellar Objects on Timescales of Days to Years}

\correspondingauthor{Jeong-Eun Lee}
\email{lee.jeongeun@snu.ac.kr}

\author{Sieun Lee}
\affil{School of Space Research, Kyung Hee University,
1732, Deogyeong-daero, Giheung-gu, Yongin-si, Gyeonggi-do 17104, Korea}

\author[0000-0003-3119-2087]{Jeong-Eun Lee}
\affil{Department of Physics and Astronomy, Seoul National University, 1 Gwanak-ro, Gwanak-gu, Seoul 08826, Korea}
\affil{SNU Astronomy Research Center, Seoul National University, 1 Gwanak-ro, Gwanak-gu, Seoul 08826, Republic of Korea}

\author[0000-0003-1894-1880]{Carlos Contreras Pe{\~n}a}
\affil{Department of Physics and Astronomy, Seoul National University, 1 Gwanak-ro, Gwanak-gu, Seoul 08826, Korea}

\author[0000-0002-6773-459X]{Doug Johnstone}
\affil{NRC Herzberg Astronomy and Astrophysics, 5071 West Saanich Road, Victoria, BC, V9E 2E7, Canada}
\affil{Department of Physics and Astronomy, University of Victoria, 3800 Finnerty Road, Elliot Building, Victoria, BC, V8P 5C2, Canada}

\author{Gregory Herczeg}
\affil{Kavli Institute for Astronomy and Astrophysics, Peking University, Yiheyuan 5, Haidian Qu, 100871 Beijing, China}

\author[0000-0001-6324-8482]{Seonjae Lee}
\affil{Department of Physics and Astronomy, Seoul National University, 1 Gwanak-ro, Gwanak-gu, Seoul 08826, Korea}

\begin{abstract}
Variability in the brightness of Young Stellar Objects (YSOs) is a common phenomenon that can be caused by changes in various factors, including accretion, extinction, disk morphology, interactions between the disk and the stellar photosphere, and the rotation of hot or cold magnetic spots on the stellar photosphere. Analyzing the variability on different timescales gives insight into the mechanisms driving the brightness changes in YSOs. We investigate the variability of YSOs on both long and short timescales using two mid-infrared datasets: the NEOWISE 7.5-year W2 (4.6$\mu$m) data and the YSOVAR 40-day Spitzer/IRAC2 (4.5$\mu$m) data, respectively. We classify the variability types in each timescale following \citet{park21}. We find a higher detection rate of variable sources in the short-term (77.6\%) compared to the long-term (43.0\%) due to the higher sensitivity of the Spitzer observations. 
In addition, the higher cadence of the YSOVAR data results in the weeks-long short-term variability being mostly secular, while the years-long long-term variability explored with the coarsely sampled NEOWISE data is mostly stochastic.
By cross-matching the two catalogs, we statistically analyze the variability types exhibited by YSOs across both timescales. 
The long-term variability amplitude is mostly three times (up to ten times) greater than the short-term variability.
Furthermore, we evaluate variability on very short (1--2 day) timescales and recover a trend of the increasing amplitude of variability as the timescales increase. By comprehensively analyzing the variability of YSOs over various timescales, we contribute to a deeper understanding of the underlying mechanisms driving their variability.
\end{abstract}


\section{Introduction} \label{sec:1}

Variability is a common characteristic observed in Young Stellar Objects (YSOs). Extensive time-series observations have been conducted in regions of star formation to investigate the underlying physical mechanisms driving the variability of these systems. These studies have spanned various timescales and have covered a wide range of wavelengths, including optical, near-infrared, mid-infrared, and sub-mm observations \citep[e.g.,][]{1994Herbst, 2001Carpenter, morales11, 2021LeeYH}. The observed changes in the brightness of YSOs can be attributed to several mechanisms that affect the star, its disk, and its envelope. The physical cause of these changes includes rotational modulation of cold or hot spots on the stellar surface \citep{1992Attridge, 1996Choi}, unstable accretion processes \citep{2013Romanova, 2014Stauffer}, variations in extinction due to surrounding material \citep{2013Bouvier}, disk instabilities \citep{2023Bino}, and the presence of eclipsing binary systems \citep{2004Stassun, 2006Stassun, 2008Cargile}.

Time-series photometry at mid-infrared (mid-IR) or longer wavelengths is a valuable tool for studying the properties of the young protostars that are still deeply embedded within their nascent envelopes. A recent study by \citet{park21}, hereafter referred to as \citetalias{park21}, focused on analyzing 6000 known Young Stellar Objects (YSOs) in nearby star-forming regions using
6.5 years of mid-IR (3.4$\mu$m and 4.6$\mu$m) time-series photometry from the NEOWISE mission provided at a 6 months cadence. 
Using the averaged single-epoch exposures from NEOWISE, \citetalias{park21} discovered and characterized the long-term lightcurves of 1700 variable YSOs. Their analysis revealed that changes in extinction along the line of sight and variable accretion were the primary mechanisms driving the mid-IR variability in these YSOs. Due to the cadence of the NEOWISE epoch-averaged observations used in their study, \citetalias{park21} did not investigate variability at timescales shorter than 180 days, although several YSOs with significant variability on short (1-2 day) timescales were noted.

Short timescale (up to $\sim$ 40 days) mid-IR variability of YSOs has been extensively studied through the Young Stellar Object VARiability (YSOVAR) program \citep{morales11,2014Cody,2014Stauffer,2018Wolk}. For instance, \citet{morales11} conducted a comprehensive monitoring of YSOs in the Orion Nebula Cluster (ONC) region using high-cadence mid-IR Spitzer IRAC1 (3.6$\mu$m) and IRAC2 (4.5$\mu$m) data over a 40-day period. Their work, along with the broader YSOVAR program, demonstrated that a significant fraction of YSOs exhibits variability on day and month timescales, 
attributed to rotational modulation from stellar spots, variable extinction (analogous to AA Tau stars), and stochastic bursts \citep[see also][]{2014Cody,2014Stauffer}.

Our study aims to establish connections between YSOs' short- and long-term mid-IR variability by collating data from the {\it Spitzer} and {\it WISE} missions. Specifically, we analyze the mid-IR data of YSOs in the ONC with photometric observations available in both the YSOVAR \citep{morales11} and NEOWISE \citepalias{park21} datasets. Through this analysis, we aim to gain a deeper understanding of the underlying physical mechanisms driving YSO variability across timescales ranging from 1-2 days to up to 7.5 years ($\sim 2500$ days).

\section{DATA} \label{sec:2}
To analyze the variability of YSOs over many timescales, we analyze data from two surveys: YSOVAR and NEOWISE. Here, we provide an overview of these surveys and the methods employed for data analysis.

\subsection{NEOWISE : Long-term (years) data} \label{subsec:2.1}
The NEOWISE (Near-Earth Object WISE) survey extends the WISE mission. Since its reactivation in 2013, NEOWISE has been collecting all-sky photometric data at 3.4$\mu$m (W1) and 4.6$\mu$m (W2) \citep{2014Mainzer}. During a single epoch of observation, about 10 $\sim$ 20 exposures are taken over a span of a few days, typically ranging over 1 to 2 days. These observations are repeated every 6 months.

In \citetalias{park21}, a catalog of 5398 nearby YSOs was compiled, encompassing nearby star-forming regions such as Orion A, B \citep{megeath12}, Taurus \citep{esplin19}, and the rest of the Gould Belt \citep{dunham15}. The NEOWISE W1 and W2 photometric data acquired between 2013 and 2020 were extracted for this YSO sample. For each object, \citetalias{park21} averaged 1-2 days of exposures within each epoch to create lightcurves with a single photometric point every six months. This resulted in a maximum of 14 epochs per source, enabling the characterization of the long-term 6.5-year mid-IR photometry. The NEOWISE 2022 data release\footnote{Available at : \cite{https://doi.org/10.26131/irsa143}}
includes photometry from two additional epochs that were not available at the time of the \citetalias{park21} analysis. This new data release expands the observational baseline of NEOWISE to 7.5 years.

\begin{figure}
\centering
{\includegraphics[trim={4cm 0cm 6cm 2cm},clip,width=1\columnwidth]
{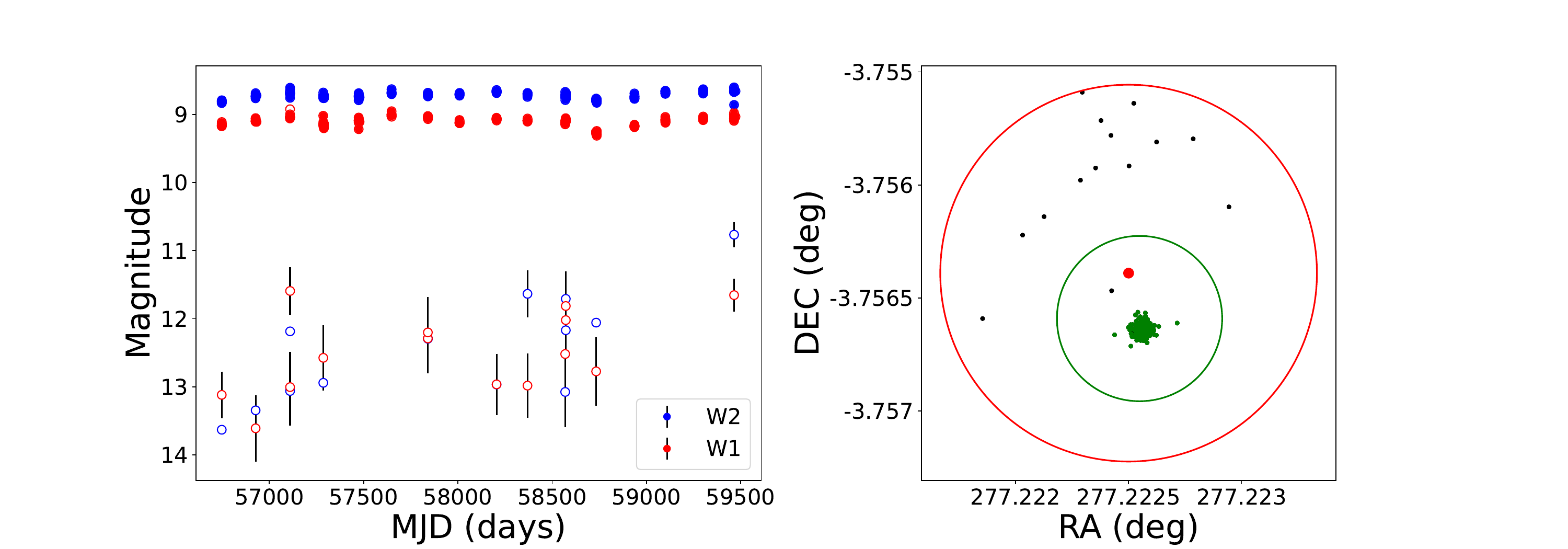}}
\caption{The excluded NEOWISE low-quality exposures. 
(Left) The plot represents the original photometric data for W1 (red) and W2 (blue). Solid circles indicate the exposures that satisfied the distance and signal-to-noise ratio criteria, while open circles represent the data points that did not meet the criteria.
(Right) The x-axis corresponds to the right ascension (RA), and the y-axis corresponds to the declination (Dec). The red dot at the center represents the coordinates extracted from the original catalog, and the red circle indicates the 3\arcsec search radius. The black dots represent all the original detections that are subsequently removed due to our criteria, the green circle represents the 2*$sd_{d}$ region, and the green dots represent the exposures that satisfy this cut.
\label{fig:exposure}}
\end{figure}

For our study, we used the single-epoch catalog of NEOWISE and performed a query within a 3\arcsec\ radius around the coordinates of each YSO, using the same star-forming region catalogs as  \citetalias{park21}. For this set of measurements, we calculated the mean and standard deviation of the distances ($sd_{d}$) between the NEOWISE single exposure coordinates and the known coordinates of the YSO. We included exposures that met the following criteria (Figure \ref{fig:exposure}):

\begin{itemize}
    \item Signal-to-noise (SNR)$ > 7$;
    \item Frame quality score (qual\_frame) $>$0;
    \item The source in the NEOWISE single exposure was located within 2*$sd_{d}$ of the mean location overall NEOWISE measurements (where $sd_{d}$ is the standard deviation).
\end{itemize}

To ensure adequate data for each investigated source, we selected targets with more than 32 NEOWISE exposures, corresponding to an average of at least 2 exposures per single epoch. Subsequently, we averaged the single-epoch exposures, typically obtained within 1 to 2 days, to generate regularly sampled lightcurves with measurements available every 6 months.

Following the approach of \citetalias{park21}, we applied additional criteria to construct a high-quality dataset to analyze YSO variability. For our analysis, the targets had to satisfy the following conditions:

\begin{itemize}
\item Detected in more than 14 epochs in W2;
\item Standard deviation of distance from the known YSO coordinates smaller than 0.3\arcsec; 
\item Mean W2 uncertainty smaller than 0.2 magnitudes. 
\end{itemize} 
After applying these criteria, we retained 5398 sources for further analysis. We note that specific criteria used in analyzing the raw NEOWISE data for our study differ from those employed in the \citetalias{park21} study. Our criteria are stricter to select only the most robust measurements.
 Here, we note that the total number of retained sources in this study and \citetalias{park21} is coincidentally the same, but individual sources are not identical.

\citetalias{park21} employed less strict criteria than ours for individual exposures. Thus, they chose to eliminate outlying data points, considering only 70\% of the data. During our analysis, however, we recognized that the excluded 30\% of the data demonstrated sufficient variability potential. As a result, we introduced the aforementioned strict conditions and retain the entire dataset, as detailed in Appendix \ref{app:1}.

\subsection{YSOVAR : Short-term (weeks) data} \label{subsec:2.2}
YSOVAR is a survey that used the {\it Spitzer} Space Telescope to investigate the mid-IR photometric variability of YSOs. The pioneering work of \cite{morales11} focused on the YSOVAR survey conducted in the ONC region, covering an area of 0.9 deg$^2$. The observations were carried out in the Fall of 2009, with time-series imaging in the mid-IR bands of IRAC1 (3.6 $\mu$m) and IRAC2 (4.5 $\mu$m). 
The observations were conducted over 40 consecutive days, with a cadence of approximately two observations per day, resulting in 81 exposures, in total, in each band. This high-cadence monitoring allowed for a detailed analysis of short-term variability of YSOs within the ONC region. 
The resulting lightcurve catalog from \cite{morales11} contained 2067 sources, which served as valuable data for investigating the short-term variability of YSOs. This dataset was used in our study to analyze the short-term mid-IR variability of YSOs and establish connections with the long-term variability observed in the NEOWISE data.

\section{YSO VARIABILITY CLASSIFICATION} \label{sec:3}
\begin{figure}[h]
\centering
{\includegraphics[trim={1cm 1cm 1cm 1cm},clip,width=1\columnwidth]
{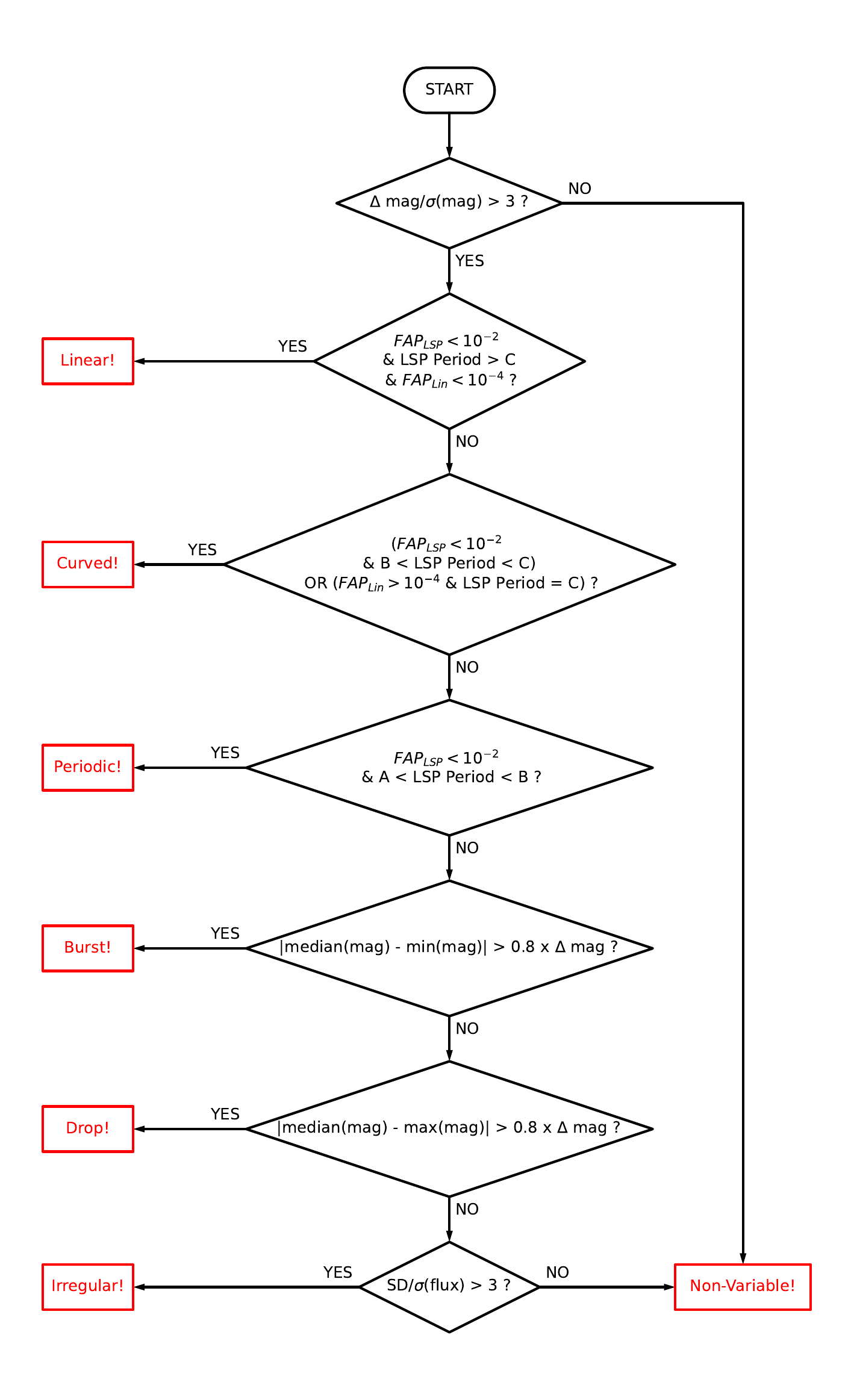}}
\caption{Variability Classification Methodology Scheme. $\Delta mag$ represents the difference between the maximum and minimum magnitudes, while $\sigma$(mag) signifies the mean value of magnitude uncertainty. $FAP_{LSP}$ denotes the False Alarm Probability from the Lomb-Scargle Periodogram, and $FAP_{Lin}$ indicates the False Alarm Probability from Linear Least Square fitting. SD stands for flux standard deviation, and $\sigma$(flux) represents the mean value of flux uncertainty. For periods, A is 200 days for NEOWISE data and 0 days for YSOVAR data. B is 1370 days for NEOWISE data and 20 days for YSOVAR data. C is 5480 days for NEOWISE data and 160 days for YSOVAR data.
\label{fig:flow_chart}}
\end{figure}

In this study, we adopted the YSO variability classification methodology proposed by \citetalias{park21} for both the 7.5-year NEOWISE data and the 40-day YSOVAR data. This classification scheme categorizes YSO variability into six distinct types: secular variables as \textit{Linear}, \textit{Curved}, and \textit{Periodic}, and stochastic variables as \textit{Burst}, \textit{Drop}, and \textit{Irregular}. The classification is based on various parameters derived from the W2 lightcurves, including the standard deviation of fluxes, the amplitude of variability, and the identification of periodicity using the Lomb-Scargle Periodogram \citep[][]{1976Lomb,1989Scargle}. Linearly rising(+) or declining(--) trends in the lightcurves are examined. Figure \ref{fig:flow_chart} depicts this variability classification methodology with a flowchart. 

We made certain modifications to the definition of secular variability (\textit{Linear}, \textit{Curved}, and \textit{Periodic} sources) compared to \citetalias{park21} in order to account for the different timescales studied using YSOVAR and the NEOWISE 2022 data release. 
The basis of the timescale used for the classification in \citetalias{park21} was the 6.5 years covered by the NEOWISE lightcurves adopted in the study.
Given the different timescales of our datasets, we adjusted the period ranges to be more appropriate for the respective data sets. The subsequent sections describe the specific values and justifications for these new period ranges.

\begin{figure*}
\centering
{\includegraphics[trim={3cm 1cm 2cm 2cm},clip,width=1\textwidth]
{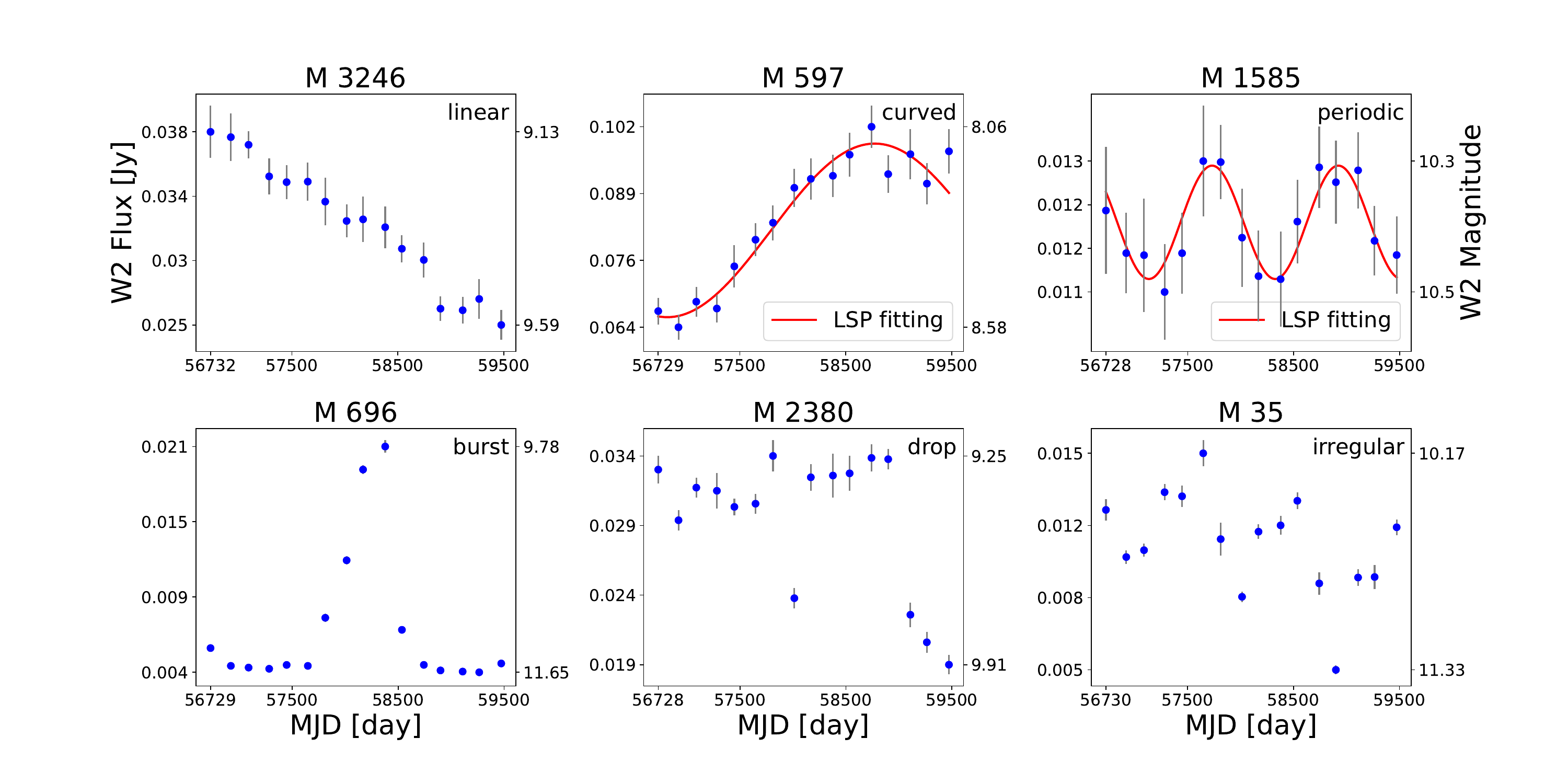}}
\caption{Examples of NEOWISE lightcurves for each long-term variability type. The source name is shown at the top of each panel, and each variable type is indicated in the upper right corner. The first observation date (MJD) is marked on the very left x-axis of each panel. \textit{Curved} and \textit{Periodic} are depicted with LSP fittings shown as red lines.
\label{fig:lc_long}}
\end{figure*}

\begin{deluxetable*}{cccccccc||c}
\tablecaption{Long-term Variable Type\label{tab:long_term}}

\tablehead{\colhead{} & \colhead{\textit{Linear}} & \colhead{\textit{Curved}} & \colhead{\textit{Periodic}} & \colhead{\textit{Burst}} & \colhead{\textit{Drop}} & \colhead{\textit{Irregular}} &
\colhead{\textit{Non-var type} }
& \colhead{Total}
}

\startdata
NEOWISE & 86 [1.6]\tablenotemark{a}& 439 [8.1]& 154 [2.9]& 114 [2.1] & 21 [0.4] & 1505 [27.9]& 3079 [57.0] & 5398 [100] \\
\hline
\textit{+Periodic}\tablenotemark{b} & - & 4 & - & - & - & - &  -& 4 (0.1) \tablenotemark{c}\\
\textit{+Burst} & - & 6 & 2 & -& -& -& -& 8 (0.1)\\
\textit{+Drop} & 3 &- &- &- &- &- & -& 3 (0.1) \\
\textit{+Irregular} &20 & 138 & 5 & -& -&- & -&163 (3.0)\\
\enddata

\tablenotetext{a}{The numbers in front are the count of variables for each type, while the numbers in the square bracket are the fractions (\%) of variables relative to their total samples. The sum of the number in square bracket is 100.}
\tablenotetext{b}{Combined variable type}

\tablenotetext{c}{The numbers in front are the total count for each combined variable type, while the numbers in parentheses are the fractions (\%) of combined variables relative to total variable samples.}
\end{deluxetable*}

\subsection{Long-term data} \label{subsec:3.1}
In our classification of long-term variability using the NEOWISE W2 observations, we followed the criteria outlined in \citetalias{park21}. The first criterion was based on the magnitude domain, specifically looking for YSOs with a magnitude difference $\Delta W2$ (${\rm = W2_{max} - W2_{min}}$) greater than three times the mean uncertainty $\sigma(W2)$ in W2 magnitudes. We identified 4218 YSOs that satisfied this condition.
For the classification of stochastic variability, we followed the same criteria defined in \citetalias{park21}.
Among the targets that satisfied the first condition, the criterion for classifying as \textit{Burst} was (median magnitude - minimum magnitude) $>$ $0.8\times$$\Delta$W2. For \textit{Drop}, the criterion was (maximum magnitude - median magnitude) $>$ $0.8\times$$\Delta$W2. For \textit{Irregular}, the criterion was SD/$\sigma>3$, where SD represents the standard deviation of flux, and $\sigma$ represents the mean flux uncertainty.

Due to the longer timescales involved in our study, we adjusted the periods used to define the different secular classes. 
YSOs with Lomb-Scargle Periodogram (LSP) periods longer than 5480 days (15 yrs) have timescales longer than twice the 7.5 yrs of NEOWISE monitoring and were therefore  classified as \textit{Linear} variables.
YSOs with periods ranging from 1370 days to 5480 days were classified as \textit{Curved} variables. Finally, sources with periods between 200 days and 1370 days were classified as \textit{Periodic} variables.

Similar to the results in \citetalias{park21}, some variable YSOs exhibit additional trends besides their dominant long-term behavior. 
We conducted tests to determine whether secular variables ($Linear$, $Curved$, $Periodic$) exhibit additional variable types. Following the flow chart (Figure \ref{fig:flow_chart}), we first removed the dominant secular trend. We then tested for secondary secular or stochastic variability by continuing the flow chart analysis, reinserting the residual light curve at a lower decision box (diamond). Thus, for primary $Linear$ and  $Curved$ variables, secondary $Periodic$ and $Stochastic$ variations were tested (by reinserting the residual light curve into the fourth diamond from the top) while for primary $Periodic$ variables only $Stochastic$ secondary variations were tested (by reinserting the residual into the fifth diamond from the top). If a source shows both trends (Secular + additional trend), it is classified as a combined variable type. This methodology is consistent with that of \citetalias{park21}.
However, we found some variables with $Linear$ or $Curved$ types that also exhibit a simultaneous periodic tendency. Therefore, we classified these as $Secular+Periodic$, a category not present in \citetalias{park21}.

Table \ref{tab:long_term} provides an overview of the long-term YSO variable type results based on the NEOWISE data. 
Out of the 4218 sources that satisfy the first criterion $\Delta \text{W2}/\sigma (\text{W2}) > 3$, 2319 were classified as having secular or stochastic variability types.
Among the variable YSOs, most were classified as \textit{Irregular} (64.9\%), indicating unpredictable and stochastic variability. Additionally, the combined variable type \textit{Secular+Irregular} was dominant among the YSOs exhibiting additional trends on top of the long-term behavior.
To visually illustrate the different long-term variable types, Figure \ref{fig:lc_long} presents example NEOWISE lightcurves for each variability type. These examples provide a glimpse into the diverse patterns and behaviors exhibited by YSOs over many years.

\subsubsection{Comparison between \citetalias{park21} and this work} \label{subsubsec:3.1.1}

This section will examine the reasons behind the changes in the number and types of variables between \citetalias{park21} and this study. In \citetalias{park21}, out of 5398 YSOs, 1734 were identified as variables, while in this study, 2319 were identified as variables, despite classifying variables by the same criteria. We attribute this increase {\bf mainly} to the following two factors:

\begin{itemize}
\item Data coverage per epoch: \citetalias{park21} used only 70\% of the data in each epoch, whereas this study used 100\% of the data points (see Appendix \ref{app:1}).
\item Time coverage: \citetalias{park21} used 6.5 years of NEOWISE data, while this study used an extended data set of 7.5 years.
\end{itemize} 

One more difference between \citetalias{park21} and this work is the criterion for detected epochs during initial sample determination, i.e., 5 epochs versus 14 epochs. However, the stricter criterion in this work is expected to reduce the overall number of samples and, consequently, the number of variables, which is the opposite of what was found. To examine the difference, we investigate the selection criterion of detection epochs under identical conditions (7.5 years time coverage and 100\% data coverage). According to our examination, the number of targets detected in 5 or more epochs was 5696, with 2446 variables. In contrast, the number of targets detected in 14 or more epochs was 5398, with 2319 variables, indicating a 5\% decrease in variables. As expected, the detection epoch criterion does not cause the increase of variables in this work.

\begin{figure*}
\centering
{\includegraphics[trim={0cm 0cm 0cm 0cm},clip,width=1\textwidth]
{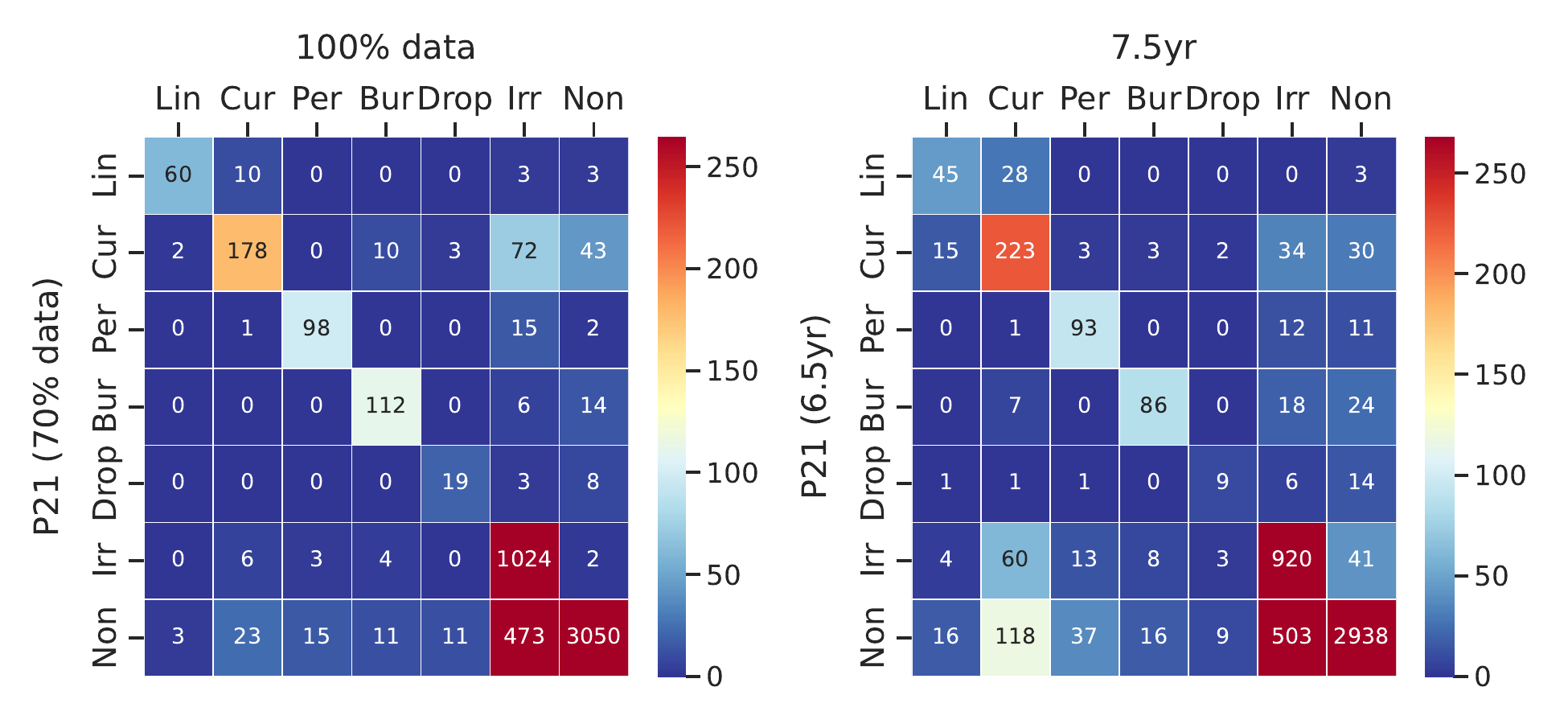}}
\caption{(Left) Comparing two analyses that used 6.5 years data but only differed in data points per epoch (70\% data (\citetalias{park21}) and 100\% data). (Right) Comparing two analyses that used 70\% data but differed in the time coverage (6.5 years and 7.5 years). 
The value within each box indicates the sources associated with that joint type.
\label{fig:heatmap}}
\end{figure*}

Therefore, we analyze the two main factors of data coverage per epoch and time coverage in more detail and compare the results with a heatmap, contrasting them with the results of \citetalias{park21} (Figure \ref{fig:heatmap}). The left panel of Figure \ref{fig:heatmap} shows the differences in classification that arise from using 6.5 years of data but analyzing 70\% (as in \citetalias{park21}) and 100\% of the data in individual epochs (this study). 
Each column represents the variability type identified when using 100\% of the data, and each row represents the variability type found when using 70\% of the data. If none of the sources changed their type between these analyses, only the diagonal boxes would have non-zero values. For instance, in the box located in the first row and second column, `10' indicates the number of sources that changed their variability type from $Linear$ when using 70\% of the data to $Curved$ when using 100\% of the data.

On the other hand, the right panel shows the differences in using 70\% of the data but varying the time coverage from 6.5 to 7.5 years. 
In this panel, each column shows the variability type found after using 7.5 years data, and each row shows the variability after using 6.5 years data. 
As a result, while \citetalias{park21} identified 1734 variables out of 5398 YSOs, changing the coverage of data points and time increased the number of variables to 2246 (left panel) and 2351 (right panel), respectively. Moreover, in both cases, there was a significant increase in the number of variables changing from \textit{Non-variable} to \textit{Irregular} types.

Therefore, in this study, which uses 100\% data points in individual epochs and extends the time coverage to 7.5 years, more variables are identified compared to \citetalias{park21}. Consequently, with the prospect of acquiring more data points with time, the number of variables will continue to increase.

\subsection{Short-term data} \label{subsec:3.2}
To classify the short-term YSO variability, we followed a similar procedure as described above and applied the methods of \citetalias{park21} to the YSOVAR IRAC2 catalog presented by \citet{morales11}. Given the 40-day timescale of the YSOVAR observations, we established LSP periods to differentiate the secular variables in the short-term. YSOs with LSP periods longer than 160 days were classified as \textit{Linear}, those with periods between 20 days and 160 days were classified as \textit{Curved}, and those with periods shorter than 20 days were classified as \textit{Periodic}.

\begin{figure*}
\centering
{\includegraphics[trim={3cm 1cm 2cm 2cm},clip,width=1\textwidth]
{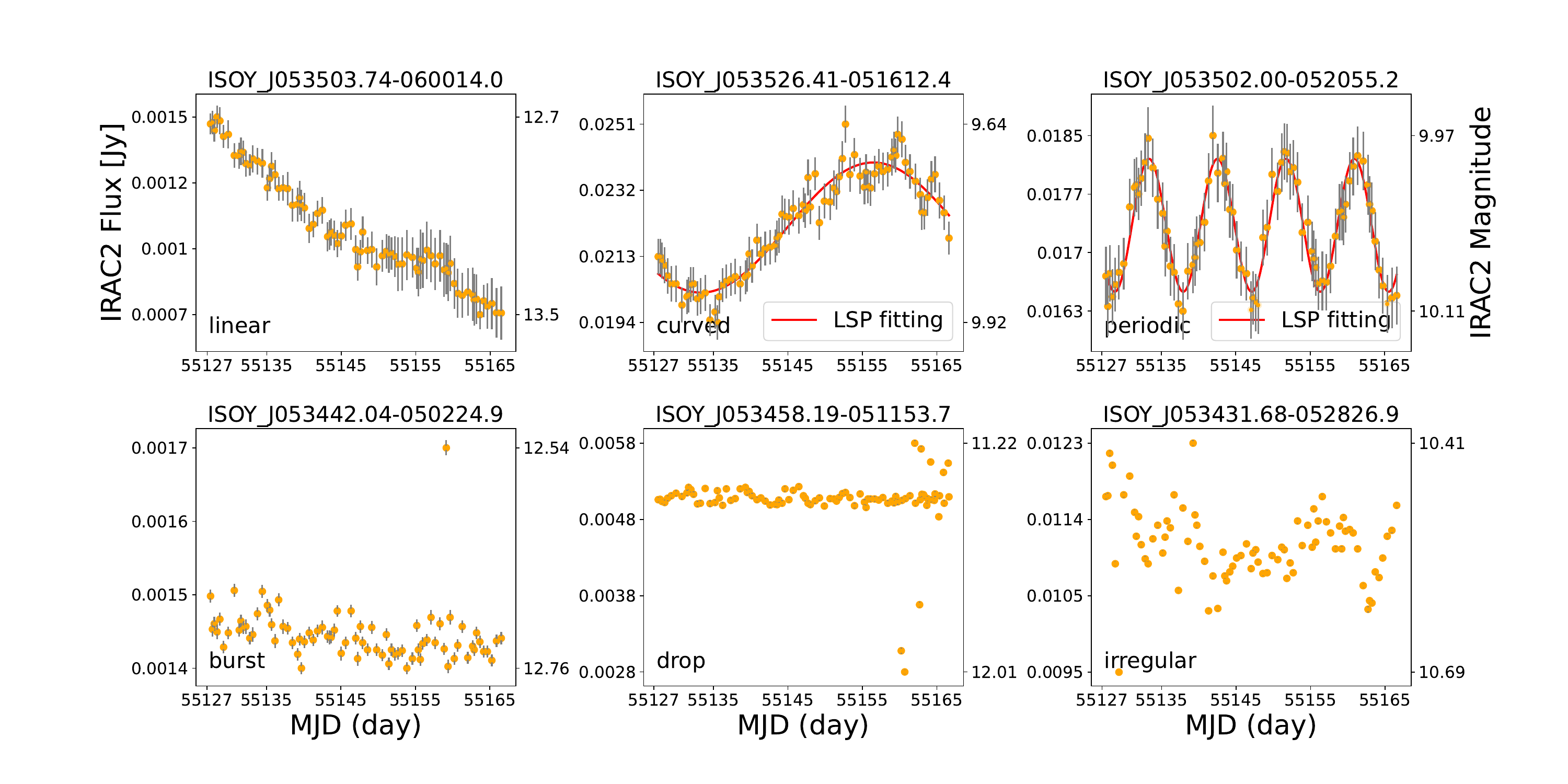}}
\caption{Examples of YSOVAR lightcurves for each short-term variability type. The source name is shown at the top of each panel, and each variable type is indicated in the bottom left corner. \textit{Curved} and \textit{Periodic} are depicted with LSP fittings shown as red lines.
\label{fig:lc_short}}
\end{figure*}

Table \ref{tab:short_term} provides an overview of the results obtained from the classification of short-term YSO variability. Similar to the long-term classification, we employed the first criterion for short-term classification, which requires $\Delta \text{IRAC2}/\sigma (\text{IRAC2}) > 3$ in the magnitude domain. Here, $\Delta \text{IRAC2}$ represents the difference between the maximum and minimum magnitudes, and $\sigma(\text{IRAC2})$ represents the mean uncertainty of the IRAC2 magnitudes. Out of the 2067 YSOVAR sources, 1807 satisfied this first criterion, and among these, 1603 YSOs (corresponding to 77.6\% of the total) were classified into different variability types. Among the variable sources, the majority were categorized as secular variables (85.5\%), with the \textit{Curved} type (49.5\%) being the most dominant among the YSO variability types. Furthermore, a significant portion of the objects classified as secular variables exhibited additional trends in their lightcurves, resulting in classification as combined variables. Additional trends were observed in 67.6\%, 70\%, and 53.3\% of the \textit{Linear}, \textit{Curved}, and \textit{Periodic} YSOs, respectively. Figure \ref{fig:lc_short} displays example YSOVAR lightcurves for each short-term variable type. These examples provide a visual representation of the different patterns and behaviors exhibited by YSOs over many weeks.

\begin{deluxetable*}{cccccccc||c}
\tablecaption{Short-term Variable Type\label{tab:short_term}}
\tablenum{2}

\tablehead{\colhead{} & \colhead{\textit{Linear}} & \colhead{\textit{Curved}} & \colhead{\textit{Periodic}} & \colhead{\textit{Burst}} & \colhead{\textit{Drop}} & 
\colhead{\textit{Irregular}} 
& \colhead{\textit{Non-var type} }
 & \colhead{Total}
}
\startdata
YSOVAR & 185 [9.0]\tablenotemark{a}& 793 [38.4]& 392 [24.5]& 5 [0.2] & 13 [0.6] & 215 [10.4]& 464 [22.4]& 2067 [100] \\
\hline\hline
\textit{+Periodic}\tablenotemark{b} & 64  & 357 & - & - & - & - & -& 421 (20.4)\tablenotemark{c}\\
\textit{+Burst} & - & 1 & 1 & -& -& - & - & 2 (0.1)\\
\textit{+Drop} &- &2 &1 &- &- &- & - &3 (0.1)\\
\textit{+Irregular} &61 & 195 &207 & -& -&- & - &463 (22.4)\\
\enddata
\tablenotetext{*}{The table note mark is same as Table 1.}
\end{deluxetable*}

\subsubsection{Comparison between YSOVAR and this work} \label{subsubsec:3.2.1}

This section discusses the differences between the variability analysis conducted in the YSOVAR study and the present study.

In \citet{morales11}, the authors classified the variability in their targets using the Stetson index \citep{1993Welch} to search for correlated variability in both IRAC channels. They also included YSOs with determined periods, even if they did not exhibit high values of the Stetson index. The classification into different types was supported by visual inspection of the lightcurves and complemented by contemporaneous photometry at shorter near-IR wavelengths. The variability in their sample was attributed to physical mechanisms such as photospheric spots, changes in disk geometry, variable extinction (AA Tau-like), and slow changes in the accretion rate.

\begin{figure*}
\centering
{\includegraphics[trim={3cm 1.5cm 2.7cm 3cm},clip,width=1\textwidth]
{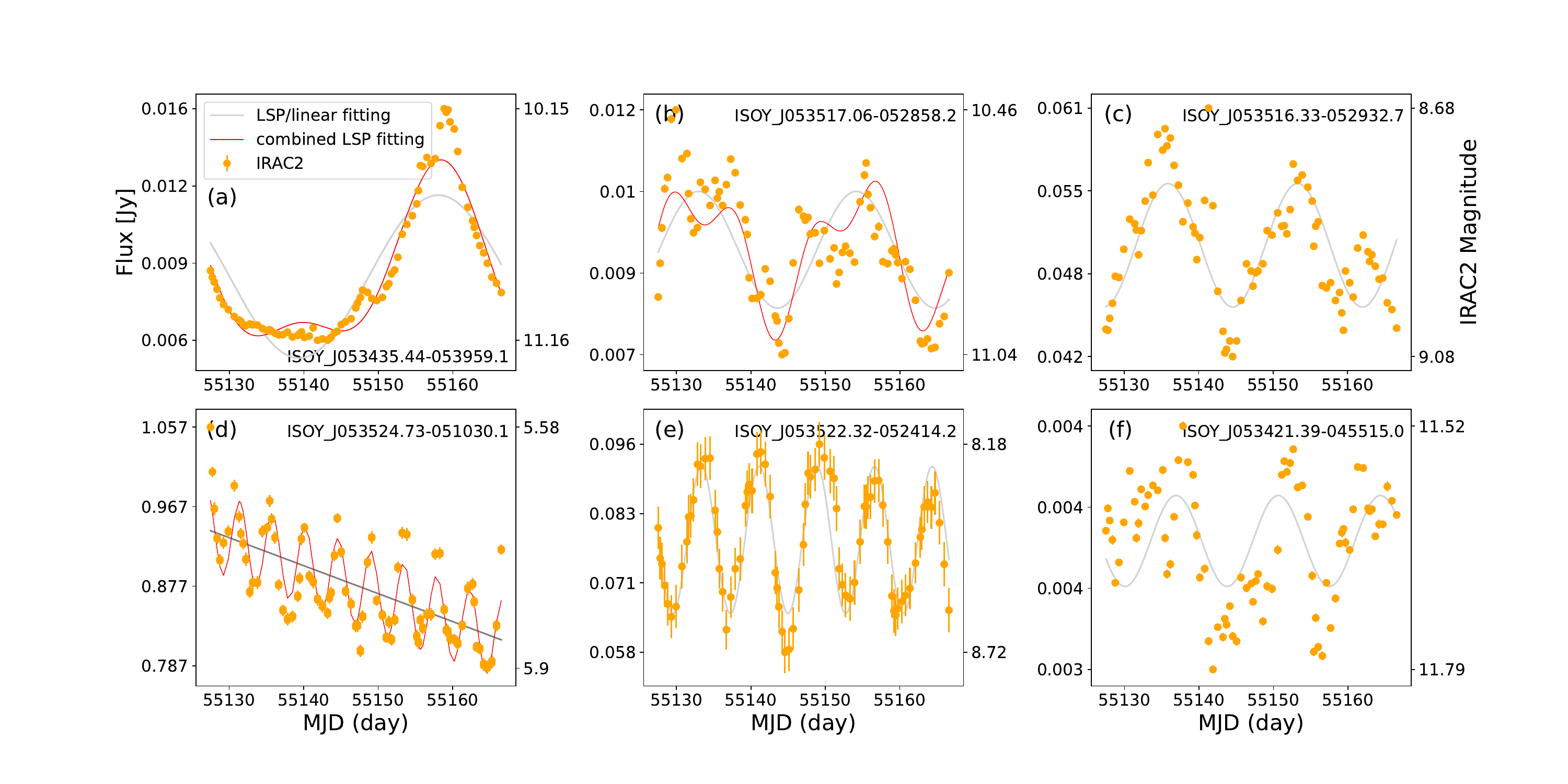} }
\caption{Examples of YSOVAR lightcurves with their LSP fits. The gray line represents the dominant variable trend while the red line includes a shorter variable trend if such a period exists. These sources are the same as in Figure 1 by \cite{morales11}. (a) and (b) Curved(gray line) + Periodic sources (red line); (c) Periodic+Irregular source; (d) Linear+Periodic source; (e) Periodic source; (f) Periodic + Irregular source.
\label{fig:yvarcomprison}}
\end{figure*}

In our study, we performed a statistical classification of the lightcurves without relying on visual inspection. Our method aims to describe the shape of the lightcurve without immediately associating the behavior with a specific physical mechanism. Additionally, we did not use additional photometry at shorter wavelengths. Figure \ref{fig:yvarcomprison} displays the fits and classification for six sources from Figure 1 by \citet{morales11}, which were used to illustrate the different types of variability identified in their work. Based on our classification alone, it is impossible to discern between the different mechanisms proposed by \citet{morales11}. However, all six YSOs were classified as \textit{Periodic} (or \textit{Periodic+Stochastic}) in our analysis, indicating variability at timescales shorter than 20 days, which is likely associated with the phenomena in the stellar photosphere and inner disk, as proposed by \citet{morales11}. Furthermore, we note that the longer timescale trends observed in the YSOVAR study are revealed by the \textit{Linear} classification in our analysis.

\begin{figure*}
\centering
{\includegraphics[trim={3cm 1.5cm 2.7cm 3cm},clip,width=1\textwidth]
{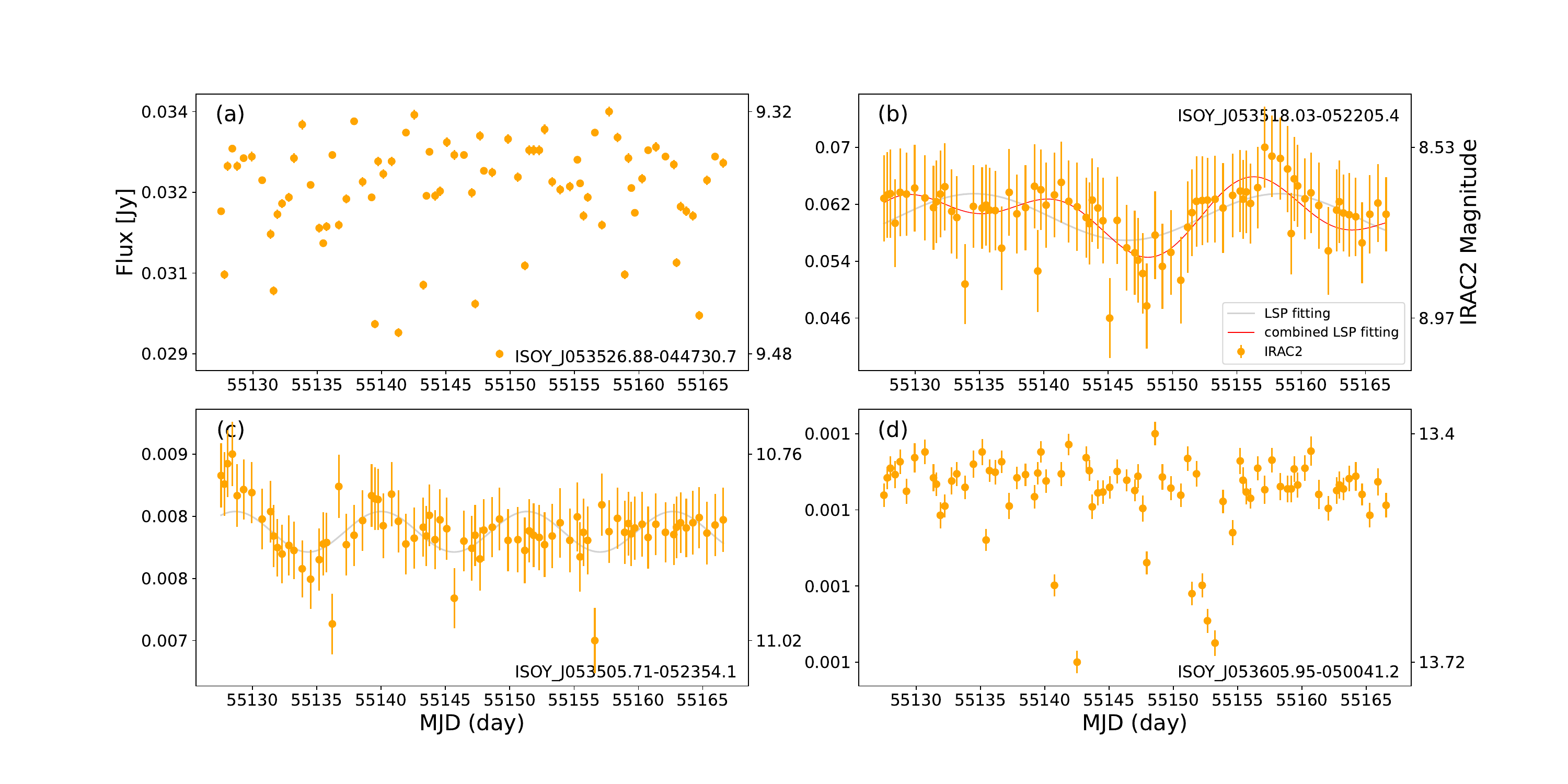} }
\caption{Candidate PMS eclipsing binary lightcurves identified by \cite{morales11}. These are sources listed in Table 5 of \cite{morales11} where they identified a stochastically repeating variability. We are unable to reproduce their results by our analysis. Using our criteria, these sources are classified as (a) Irregular, (b) Curved+Periodic, (c) Periodic, and (d) Drop
\label{fig:yvarcomprisont_t5}}
\end{figure*}

Regarding the identification of PMS eclipsing binaries, \citet{morales11} reported nine cases, four of which were newly detected systems. However, when applying our methods, we find it more challenging to distinguish these four candidate PMS binaries compared to their results. Figure \ref{fig:yvarcomprisont_t5} illustrates that we are only able to classify one of their candidate YSOs as \textit{Periodic}, but when looking at the fitting, it is difficult to consider it as an eclipsing binary.

\begin{figure}[ht!]
\centering
{\includegraphics[trim={0.3cm 0.1cm 0.1cm 0.0cm},clip,width=0.98\columnwidth]
{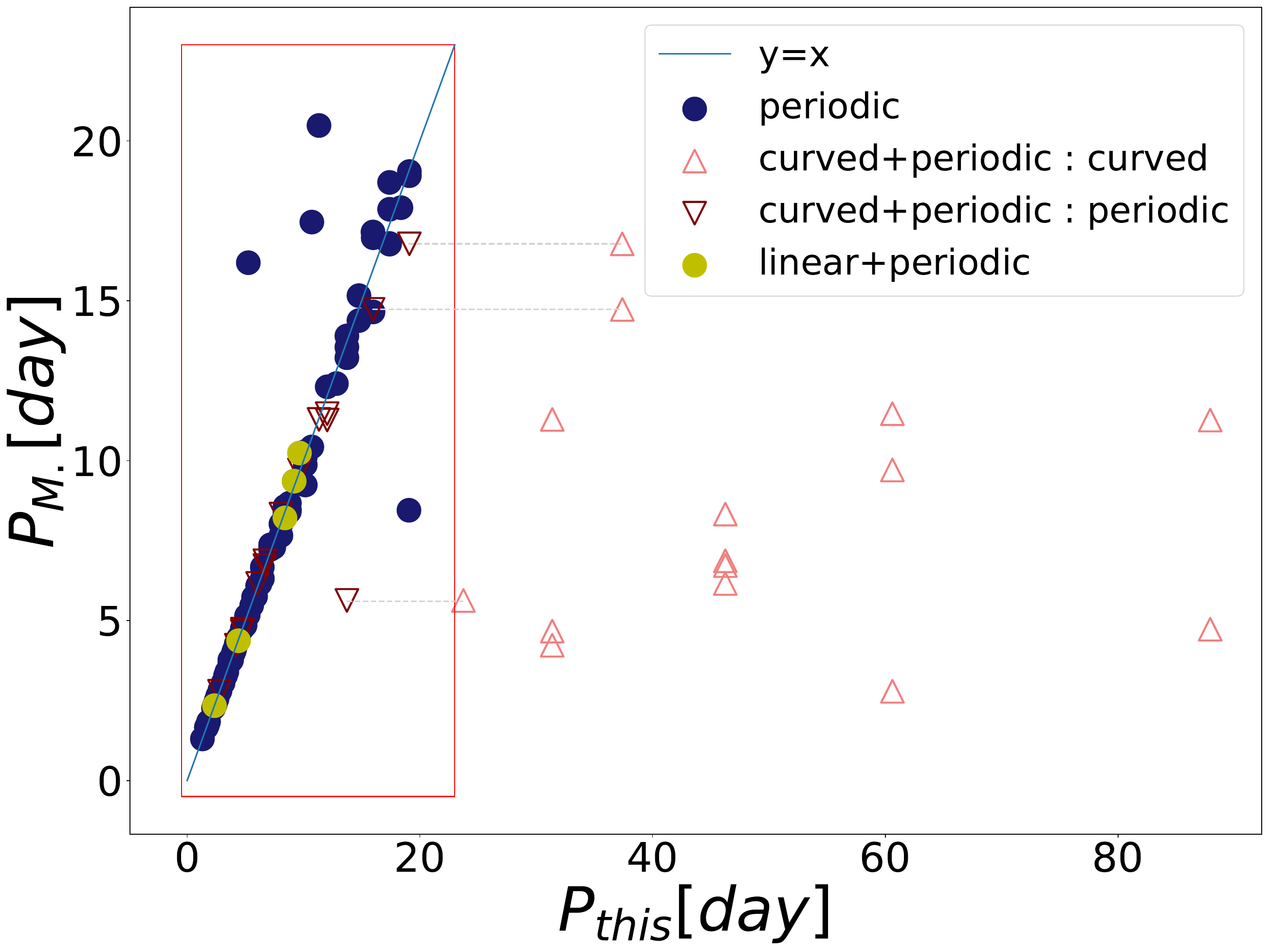} }
\caption{Scatter plot comparing the periods extracted from the YSOVAR data by the previous and current studies. The x-axis represents the periods calculated in this study, while the y-axis represents the periods calculated by \cite{morales11}. The plotted data points represent sources classified as \textit{Periodic} or \textit{+Periodic} as a combined variable. 
For sources classified as \textit{Curved}+\textit{Periodic} (15 sources), which exhibit two distinct periods corresponding to the \textit{Curved} classification and the +\textit{Periodic} classification, they are depicted by pink triangles and brown upside-down triangles, respectively. The two periods are connected by gray dotted lines only for three sources as examples.
Sources classified solely as \textit{Periodic} (84 sources) are denoted by blue dots, while the +\textit{Periodic} periods of sources classified as \textit{Linear}+\textit{Periodic} are represented by olive dots (5 sources). The blue line represents the one-to-one relation (y=x).
\label{fig:yvarperiod}}
\end{figure}

When comparing all the periodic variables, however, the analysis of the previous study and the present study yield very similar results. As shown in Figure \ref{fig:yvarperiod}, the periods derived using our method align well with the periods determined by \cite{morales11}, except for five out of 104 sources. Most periods lie along the one-to-one relation (blue line). We also indicate objects that exhibit two periods in our analysis, classified as \textit{Curved+Periodic}. This represents one of the main differences from the work by \citet{morales11}, as these additional trends were not identified in their study.

\section{Relating SHORT- AND LONG-TERM VARIABILITY} \label{sec:4}

To study the connection between short-  and long-term, weeks and years, variability in YSOs, we crossmatched the two catalogs described in the previous sections.

\subsection{Cross-matched sources} \label{subsec:4.1}
We conducted a cross-match between the NEOWISE (Section \ref{subsec:3.1}) and YSOVAR (Section \ref{subsec:3.2}) catalogs using a matching radius of three arcseconds, resulting in 703 positive matches. Table \ref{tab:table3} provides the number of these YSOs classified into different variability types based on both the short- and long-term analyses.
A significant portion of the sources (58.2\%) do not have a NEOWISE (long-term) variability classification. In contrast, the YSOVAR (short-term) classifications reveal that the majority of sources (93.0\%) are classified as variable YSOs. This discrepancy can be attributed to the higher sensitivity of the {\it Spitzer} telescope compared to the {\it WISE} telescope enabling the detection of variability types with lower amplitudes. For the cross-matched sources, the median magnitude uncertainty was 0.005 in YSOVAR and 0.04 in NEOWISE, indicating that the sensitivity of YSOVAR is approximately 8 times better than that of NEOWISE.
Some YSOs likely exhibit high-frequency and low-amplitude variability that remains relatively stable over time, without additional mechanisms causing significant flux changes over longer observation periods. As such, NEOWISE, with its poorer sensitivity, would not classify these YSOs as variable stars in the long-term, whereas the YSOVAR data would still be capable of capturing the variability in these systems.

The higher cadence of the YSOVAR observations provides an opportunity to study the complexity of YSO variability in more detail. In the short-term analysis, a higher percentage of YSOs are classified as combined variables compared to the long-term analysis (see Table \ref{tab:table4}). Out of the 703 sources, 509 (72.4\%) are classified as combined variables, with the majority of them falling into the \textit{Curved + Periodic} class. On the other hand, in the long-term data, only 21 sources (3.0\%) exhibit combined variable types, and they are predominantly classified as \textit{Curved + Irregular}.

We note that the Orion YSO catalog adopted by \citetalias{park21} lacks the class of PMS+E, which denotes the evolutionary stage of Class III, possibly contaminated by evolved stars. As a result, in the cross-matched sources (Table 3), the `{\it periodic}' variables almost disappear; only 0.4\% of cross-matched sources are classified as {\it periodic} in long-term. The fraction of {\it periodic} sources classified from all the NEOWISE sources is 2.9\% as shown in Table 1 (refer to \citetalias{park21} for additional discussion of the periodic classification).

\begin{deluxetable*}{cccccccc||c}
\tablecaption{Variable Type : Cross-matched Sources\label{tab:table3}}
\tablenum{3}

\tablehead{\colhead{} & \colhead{Y.\textit{Linear}\tablenotemark{a}} & \colhead{Y.\textit{Curved}} & \colhead{Y.\textit{Periodic}} & \colhead{Y.\textit{Burst}} & \colhead{Y.\textit{Drop}} & \colhead{Y.\textit{Irregular}} &
\colhead{Y.\textit{Non-variable}} &
\colhead{NEOWISE Total}
}

\startdata
N.\textit{Linear}\tablenotemark{b} & 3\tablenotemark{d} & 3 & 2 & - & - & 1 & - & 9 [1.3]\tablenotemark{c} \\
N.\textit{Curved} &3 & 36 & 10 & - & - &5 & 1 & 55 [7.8] \\
N.\textit{Periodic} & - & 2 & 1 & - & - & - & - & 3 [0.4] \\ 
N.\textit{Burst} & 7 & 4 & - & - & - & 4 & 1 & 16 [2.3] \\ 
N.\textit{Drop} & - & 3 & 1 & - & - & - & - & 4 [0.6] \\ 
N.\textit{Irregular} & 35 & 112 & 53 & - & -& 5 & 2 & 207 [29.4] \\
N.\textit{Non-variable} & 37 & 194 & 84 & - & 1 & 48 & 45 & 409 [58.2] \\
\hline \hline
YSOVAR Total &85 [12.1] & 354 [50.4] & 151 [21.5] & 0 [0.0] & 1 [0.1] & 63 [9.0] & 49 [7.0] & 703 [100]
\enddata

\tablenotetext{a}{Variable types. Y. refers to YSOVAR variable types}
\tablenotetext{b}{Variable types. N. refers to NEOWISE variable types}
\tablenotetext{c}{The numbers in front are the count of total variables for each type, while the numbers in the square bracket are the fractions (\%) of variables relative to their total samples.}

\tablenotetext{d}{The number of sources having variable type \textit{linear} in both YSOVAR(short-term) and NEOWISE(long-term).}

\end{deluxetable*}

\begin{deluxetable*}{cccc|c||ccc|c}
\tablecaption{Combined Variable Type : Cross-matched Sources\label{tab:table4}}
\tablenum{4}
\tablehead{ & & \colhead{\textit{YSOVAR}} & &
 &
 & \colhead{\textit{NEOWISE}} & &}
\vline 

\startdata
 & Y.\textit{Linear} & Y.\textit{Curved} & Y.\textit{Periodic} & Y.Total & N.\textit{Linear} & N.\textit{Curved} & N.\textit{Periodic} & N.Total\\ \hline
\textit{$+$Periodic} & 39 & 190 & - & 229 [45.0]  & - & 1 & - & 1 [4.8]\\ 
\textit{$+$Burst} & - & - & 1 &  1 [0.2] & - &  1& -  &1 [4.8]\\
\textit{$+$Drop} & - & - & 1 & 1 [0.2]& 1 &  - &  - & 1 [4.8] \\ 
\textit{$+$Irregular} &  37 & 115  & 126 & 278 [54.5] & 4 & 14 & - &  18 [85.7]\\
\hline \hline
Total & 76 [14.9] & 305 [59.9] & 128 [25.1] & 509 (72.4)\tablenotemark{a} & 5 [23.8] & 16 [76.2] &0 [0.0] & 21 (3.0)\tablenotemark{a}
\enddata

\tablenotetext{a}{The numbers in front are the total count of combined variables, while the numbers in the parentheses are the fractions (\%) of combined variables
relative to their total samples.}
\end{deluxetable*}

\subsection{Comparison of 
the variabilities} \label{subsec:4.2}

\begin{figure}
\centering
{
\includegraphics[trim={0cm 2cm 0cm 2cm},clip,width=\columnwidth]
{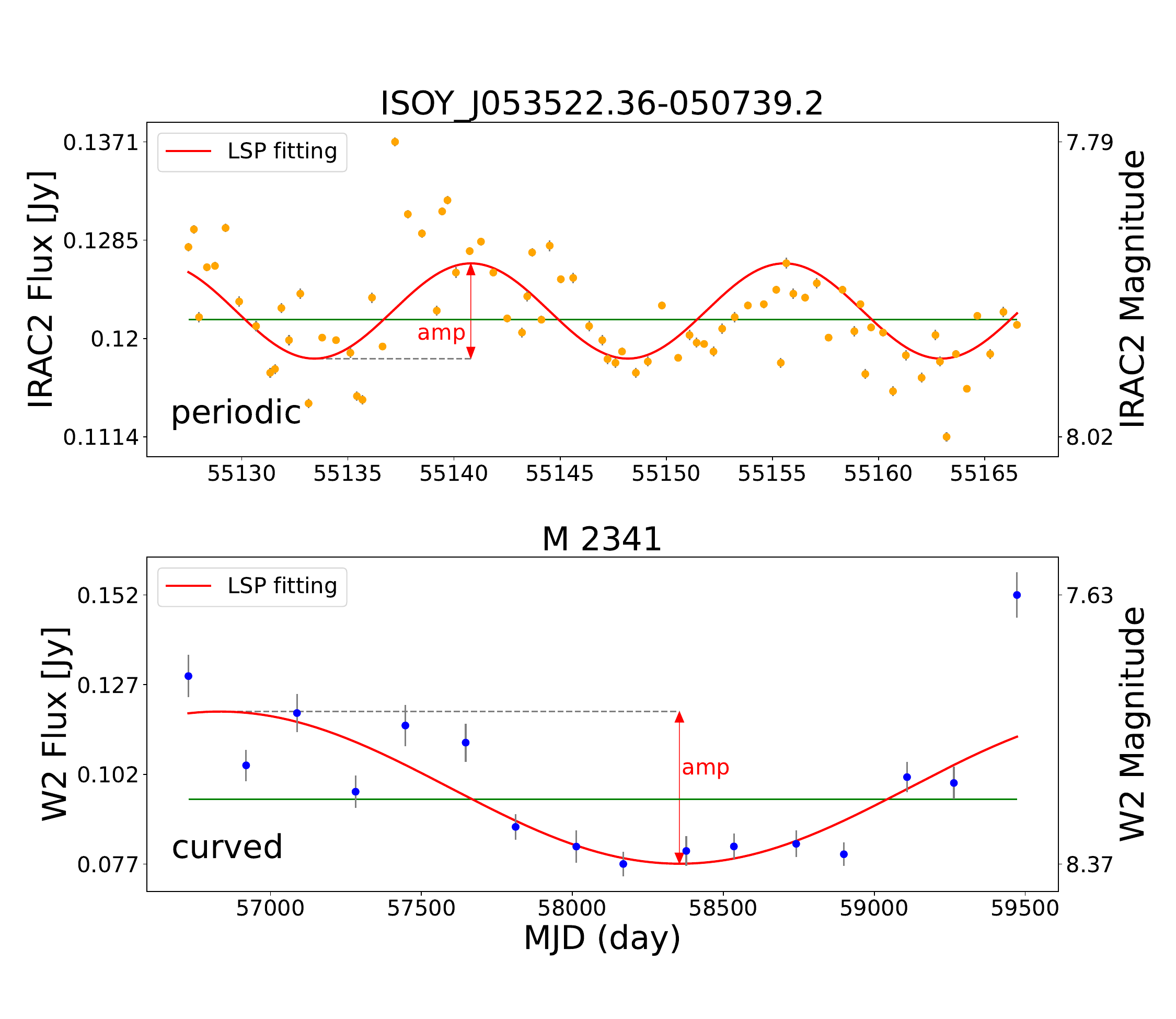} }
\caption{An example of calculating $N_{amp}$ for the YSO source ISOY\_J053522.36-050739.2 (also known as 'M2341' in the NEOWISE catalog). The top panel shows the YSOVAR lightcurve, while the bottom panel displays the NEOWISE lightcurve. The red curved lines represent the Lomb-Scargle periodogram fitting for each dataset, with the corresponding amplitudes indicated by red arrows. The green horizontal lines represent the mean flux of the fitting in each dataset. The $N_{amp}$ values were calculated as the ratio of the LSP amplitude to the mean flux of the fitting.
\label{fig:amplitude_ex}}
\end{figure}

To compare the degree of variability between the short-term and long-term observations, we analyzed two parameters: the Normalized Lomb-Scargle Periodogram amplitude ($N_{amp}$) and the normalized flux root-mean-square value ($Nf_{rms}$).
To calculate $N_{amp}$, we selected the 49 YSO sources that were classified as either \textit{Curved} or \textit{Periodic} in both the short-term and long-term analyses. For each of these sources, we obtained the long-term and short-term amplitudes from the Lomb-Scargle periodogram fitting and normalized them by dividing them by the mean flux determined during the fitting (see Figure \ref{fig:amplitude_ex}). 

\begin{figure*}
\centering
{
\includegraphics[trim={0cm 0.2cm 0cm 0.1cm},clip,width=1\textwidth]
{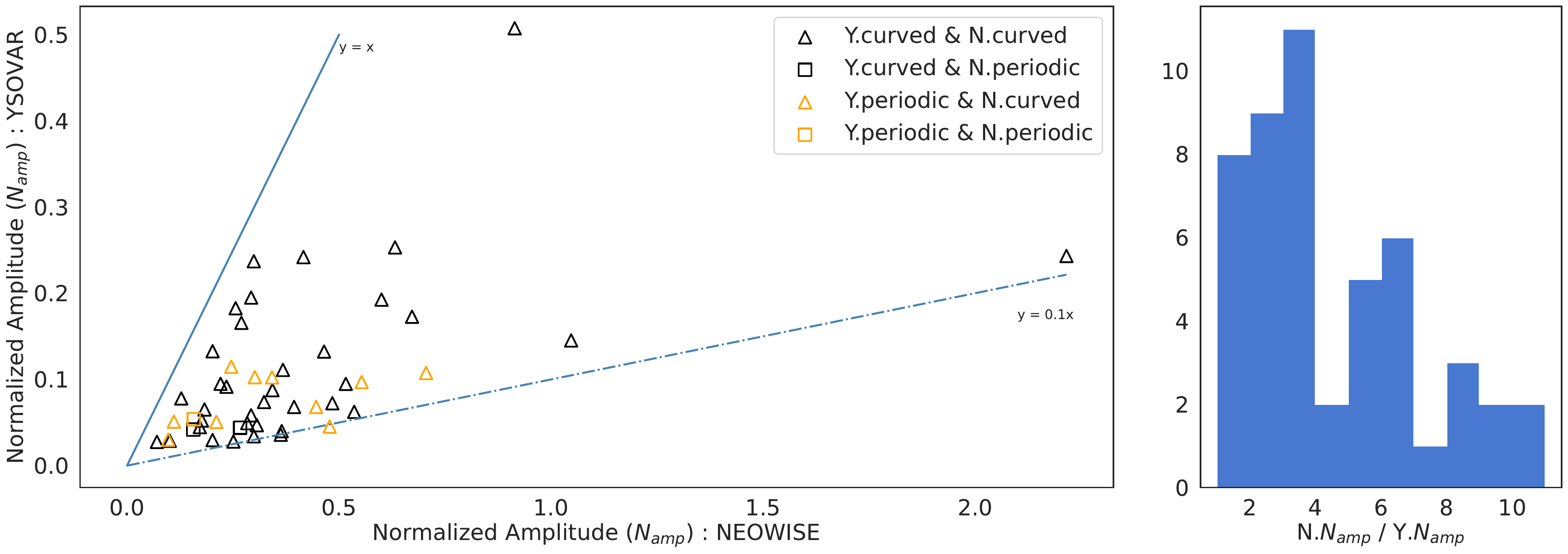} }
\caption{(Left) Scatter plot of Normalized Amplitude ($N_{amp}$) for 49 sources classified as \textit{Curved} or \textit{Periodic} in both long-term and short-term analyses. In the legend, "Y." denotes YSOVAR, and "N." denotes NEOWISE. For instance, "Y.curved \& N.periodic" signifies that a source is classified as \textit{Curved} in YSOVAR data and \textit{Periodic} in NEOWISE data. (Right) Histogram illustrating the ratio of NEOWISE normalized amplitude to YSOVAR normalized amplitude. All values are greater than 1, indicating that long-term amplitude exceeds short-term amplitude. The peak is centered around 3, and the distribution extends up to 10.
\label{fig:amplitude}}
\end{figure*}

The comparison of $N_{amp}$ between the short-term and long-term is presented in Figure \ref{fig:amplitude}. The scatter plot illustrates that the sources tend to lie to the right of the $y=x$ line, indicating that the amplitude of long-term variability is generally larger than that of short-term variability.
The histogram (right of Figure 10) shows the distribution of the amplitude ratio between NEOWISE and YSOVAR. All sources have ratios above 1, clearly signifying that long-term amplitude exceeds short-term amplitude, with a peak at $\sim$3. The largest ratio goes up to 10. This difference in amplitude can be attributed to the different underlying mechanisms driving the short-term and long-term changes. Short-term variability is primarily influenced by physical mechanisms related to the stellar photosphere or inner disk, which typically result in lower amplitudes. On the other hand, long-term changes are often driven by mechanisms involving variations in geometry and accretion rate, which can produce larger amplitude variations.

\begin{figure*}
\centering
{
\includegraphics[trim={0cm 0cm 0cm 0cm},clip,width=2\columnwidth]
{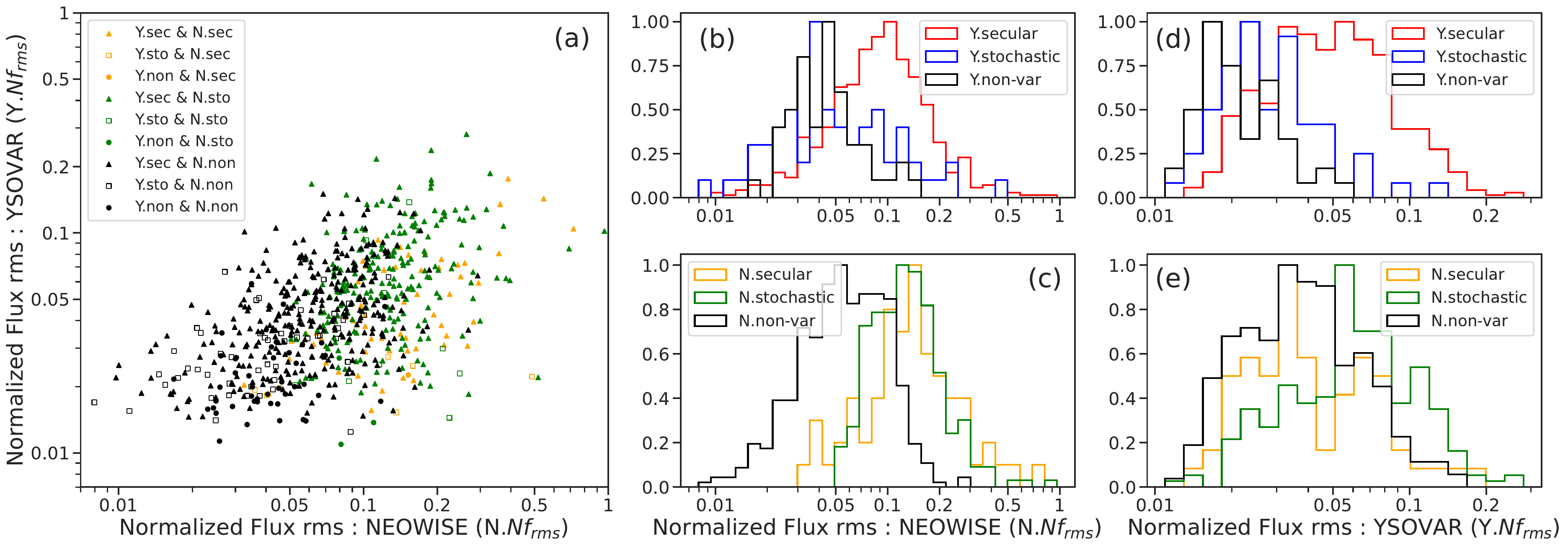} }
\caption{(a) Scatter plot depicting the Normalized Flux rms ($Nf_{rms}$) for all 703 cross-matched sources. Here, 'Y.' denotes YSOVAR, and 'N.' represents NEOWISE. The 'sec' category encompasses secular variables
, while 'sto' corresponds to stochastic variables.
'non' signifies the Non-variable type. (b) and (c) present histograms of NEOWISE normalized flux rms. (b) illustrates the classification results based on YSOVAR, while (c) showcases the classification results based on NEOWISE. Within the region of $N.Nf_{rms} <$ 0.03, some sources that were classified as secular or stochastic in the short-term data are classified as non-variable in the long-term data. (d) and (e) display histograms of YSOVAR normalized flux rms. (d) demonstrates the classification outcomes based on YSOVAR, whereas (e) illustrates the classification outcomes based on NEOWISE. The histogram's peak locations reveal that sources classified as secular in the short-term exhibit the highest variability at 
\textbf{both of} 
these timescales.
\label{fig:fluxrms}}
\end{figure*}

In a second method, we analyzed the ``normalized flux root-mean-square value ($Nf_{rms}$)" for all 703 sources. For each source, we calculated the flux rms (root-mean-square) value from the long- and short-term lightcurves and for normalization, we divided it by the median flux. Figure \ref{fig:fluxrms} presents the scatter plot and distribution histograms of $Nf_{rms}$ for the cross-matched sources.

The leftmost panel of Figure \ref{fig:fluxrms} shows the scatter plot of the Normalized Flux rms ($Nf_{rms}$) for all 703 cross-matched sources. 
The top two histogram (panels (b) and (d)) illustrates the $Nf_{rms}$ values for YSOVAR variability types, revealing that the distribution peak of secular variables (red) is at larger values compared to stochastic (blue) and non-variable (black) YSOs. This suggests that (quasi-)periodic variability related to hot spots or obscuration events dominates among the YSO variable population during the 40-day observations of the YSOVAR program. 
Interestingly, as these same sources have the highest $Nf_{rms}$ values in the NEOWISE data (panel (b)), they are also associated with long-term (NEOWISE) secular (yellow) and stochastic (green) variability (compare with panel (c)).

In the bottom panel (panels (c) and (e)), focusing on the long-term classification, secular variables (yellow) exhibit larger $Nf_{rms}$ values in the NEOWISE data ($N.Nf_{rms}$) compared to the YSOVAR data ($Y.Nf_{rms}$). This indicates that YSOs dominated by long-term, high-amplitude variability, potentially associated with significant changes in accretion or geometrical variations in the disk, tend to have relatively constant or minimally varying brightness over a few weeks of observations. 
As further support, the long-term non-variables (black) peak at roughly the same $Nf_{rms}$ value for both data sets. Notably, the distributions of long-term secular and long-term non-varying sources are almost indistinguishable in panel (e). This is not the case for long-term stochastic sources (green). These sources exhibit less of decrease in $Nf_{rms}$ values between the NEOWISE and YSOVAR datasets and show the largest $Nf_{rms}$ values in panel (e).

As noted above, 
the distribution peaks for NEOWISE stochastic variables (green) align with the $Nf_{rms}$ values of secular variables (red) classified from YSOVAR data (upper panel). This suggests that
many of the objects displaying secular variability over 40 days of observations exhibit a consistent level of variability over years-long observations. Due to the cadence of NEOWISE, these YSOs would be observed at random epochs in their lightcurves, and their short-term variability would likely appear irregular over the long term, leading to their classification as stochastic.

Figure \ref{fig:fluxrms} also highlights the higher sensitivity of {\it Spitzer} observations. In the region where $N.Nf_{rms} <$ 0.03 (panels (b) and (c)), many YSOs classified as secular or stochastic based on short-term data are categorized as non-variable in the long-term data analysis.

\subsection{Degrading the sensitivity of YSOVAR observations} \label{subsec:4.3}
To assess the potential impact of lower NEOWISE sensitivity on comparing short- and long-term variability in YSOs, we degraded the sensitivity of the YSOVAR data to examine the effect on detecting variability. In the previous section, we established that most YSOs exhibit short-term variability and suggested that this could be attributed to the superior sensitivity of the Spitzer observations compared to WISE. To verify this hypothesis, we applied the mean uncertainty of NEOWISE W2 flux to the YSOVAR flux for each source when determining YSOs as a `variable'.

Table \ref{tab:table5} compares the use of two different sensitivities. `Y.eflux' denotes the results obtained using the original YSOVAR flux errors, while `N.eflux' represents the outcomes obtained by degrading the sensitivity of the YSOVAR data to match the NEOWISE sensitivity. When determining short-term variability using the original YSOVAR flux errors, 81.9\% of YSOs were classified as variables. However, when employing the WISE flux errors, 15.1\% were identified as variables. As expected, the increased value of $\sigma$ in the YSOVAR data fails to capture the low-amplitude variability. Therefore, the telescope's sensitivity is critical in detecting YSO variability, suggesting that the NEOWISE lightcurves miss a majority of variable YSOs with low amplitudes, even in the long-term.

\begin{deluxetable}{c||cc}
\tablecaption{Variable detection by different telescope \label{tab:table5}}
\tablenum{5}

\tablehead{\colhead{Short-term data} & \colhead{Y.eflux\tablenotemark{b}} & \colhead{N.eflux\tablenotemark{c}}
}
\startdata
Variable\tablenotemark{a} & 576 [81.9] & 106 [15.1] \\
Non variable & 127 [18.1] & 597 [84.9]  \\
\hline \hline
Total & 703 [100] & 703 [100]\tablenotemark{d}
\enddata

\tablenotetext{a}{Variable = SD/$\sigma \textgreater 3$}
\tablenotetext{b}{$\sigma$ = $\sigma$(IRAC2 flux unc) ; the mean uncertainty of IRAC2 flux.}
\tablenotetext{c}{$\sigma$ = $\sigma$(W2 flux unc)  ; the mean uncertainty of W2 flux.}
\tablenotetext{d}{The numbers in the parentheses are the
fractions (\%) of variables relative to their total 703 samples.}

\end{deluxetable}

\section{Discussion} \label{sec:5}

\subsection{Variability Mechanisms} \label{subsec:5.1}

YSO variability can arise from various processes, including accretion, extinction, changes in disk properties, interactions between the disk and stellar photosphere, and hot$/$cold magnetic spots in the stellar photosphere \citep[see e.g.,][]{1994Herbst,2001Carpenter,2014Stauffer,2017Contreras,2017Cody,park21}. Comparing the short-term and long-term variability analyses from YSOVAR and NEOWISE provides insights into the mechanisms driving brightness changes in YSOs.

Analyzing weeks-long YSOVAR data alone leads to similar conclusions as those derived by \citet{morales11}. The majority of YSOs in the ONC sample (93.0\% among cross-matched YSOs in our analysis) exhibit some degree of variability. The timescales of variability are related to processes affecting the stellar photosphere-inner disk regions, such as short-term bursts caused by magnetic instabilities and reconnection between the stellar magnetosphere and the inner disk edge or periodic obscuration events similar to AA Tau-like extinction \citep[$P<15$ days, ][]{2013Bouvier,2014Cody}. Within the ONC sample, a significant fraction of YSOs (28.4\%) are classified as secular variables in the short-term analysis, which display stochastic ($Irregular$) variability in the long-term NEOWISE data. The amplitude of short-term and long-term variability in these objects shows similar values. This suggests that the mechanisms driving variability in the short term also persist over the long-term, with the classification as \textit{Irregular} arising from the cadence of NEOWISE observations. 51.8\% of the sample exhibit short-term variability but are classified as non-variable in the long-term. These objects tend to have lower amplitudes, indicating that we are likely observing the same variability mechanisms over the long-term but cannot classify them due to the lower sensitivity of WISE observations.

\begin{figure*}[ht!]
\centering
{\includegraphics[trim={4.5cm 1cm 3cm 4cm},clip,width=1\textwidth]
{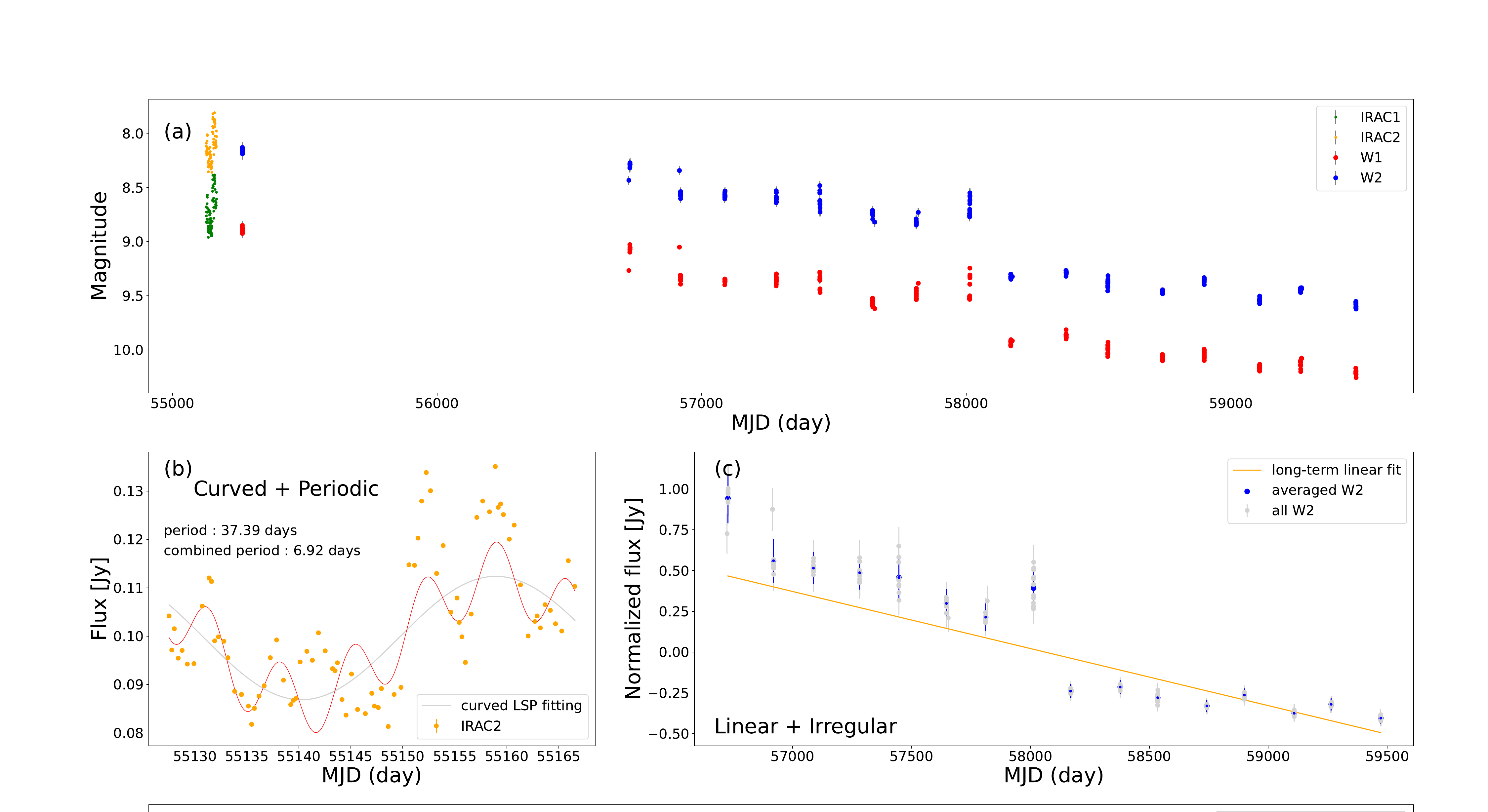}}
\caption{An example lightcurve of Disk source (V418 Ori).
(a) The overall lightcurves, including IRAC1 (green), IRAC2 (yellow), W1 (red), and W2 (blue) bands, are represented on a magnitude scale. W1 and W2 data at $\sim$ MJD 55200 are AllWISE data.
(b) The lightcurve specifically for the IRAC2 band, showing short-term variability. This source is classified as \textit{Curved+Periodic} in the short-term analysis, with periods of 37.39 days and 6.92 days. Gray and red lines represent the corresponding Lomb-Scargle Periodogram (LSP) fitting for each period, with the y-axis representing the flux scale.
(c) The W2 NEOWISE lightcurve, revealing long-term variability. This target is classified as \textit{Linear+Irregular}. The orange line represents the Linear Least Square fitting for this classification, with the y-axis representing the normalized flux, calculated as (flux - median flux) $/$ median flux.
\label{fig:veryshort1}}
\end{figure*}

YSOs classified as secular variables in the years-long NEOWISE data generally exhibit larger amplitudes in the long term compared to the short-term observations. In these cases (approximately 9.7\% of the sample), the variability may be attributed to structural changes in the disk such as changes in the scale height of the inner disk \citep{2021Covey}, long-term obscuration caused by a dusty wind \citep{2018Davies}, or variations in the accretion rate \citep{2022Fischer}. Evidence for accretion variations driving mid-IR variability comes from a direct comparison of sub-mm variability and mid-IR variability in the same source, as shown by \citet{Contreras20}.
Figure \ref{fig:veryshort1} illustrates the example of V418 Ori, a YSO displaying \textit{Curved+Periodic} short-term variability and long-term changes with higher amplitudes classified as \textit{Linear+Irregular}.

\subsection{Very short-term variability} \label{subsec:5.2}

The NEOWISE single-epoch catalog offers observations with durations mostly ranging from 1 to 2 days, repeated every 6 months (see Section \ref{subsec:2.1}). In this section, we focus on analyzing the variability at these timescales, which we refer to as the very short-term.

There are limitations in analyzing very short-term data due to its brief duration and a 6-month time gap between observations. Therefore, it becomes challenging to determine whether the observed variability in the very short-term data will persist beyond that specific time period. Additionally, due to the limited number of data points within each NEOWISE epoch, it is crucial to ensure the inclusion of only high-quality data. Consequently, we applied additional criteria to select NEOWISE high-quality data, extending beyond the considerations made in the analysis by \citetalias{park21} (see Section \ref{subsec:2.1}). With these carefully chosen high-quality exposures, we used the individual exposures, which were averaged for the long-term analysis, to examine the very short-term variability.

To investigate the behavior of very short-term variability, we visually examined the lightcurves to compare the observed variability with the expected trends derived from fitting the short- and long-term variations. In cases where a YSO was classified as \textit{Curved} or \textit{Periodic} based on the YSOVAR data (representing short-term variability) and defined as a secular variable in the NEOWISE data (representing long-term variability), we combined the YSOVAR Lomb-Scargle Periodogram (LSP) fitting with the NEOWISE long-term fitting. Subsequently, we examined whether the 1-2 day NEOWISE observations (representing very short-term variability) followed this combined fitting trend.

\begin{figure*}[ht!]
\centering
{\includegraphics[trim={3.8cm 1.9cm 3cm 1cm},clip,width=1\textwidth]
{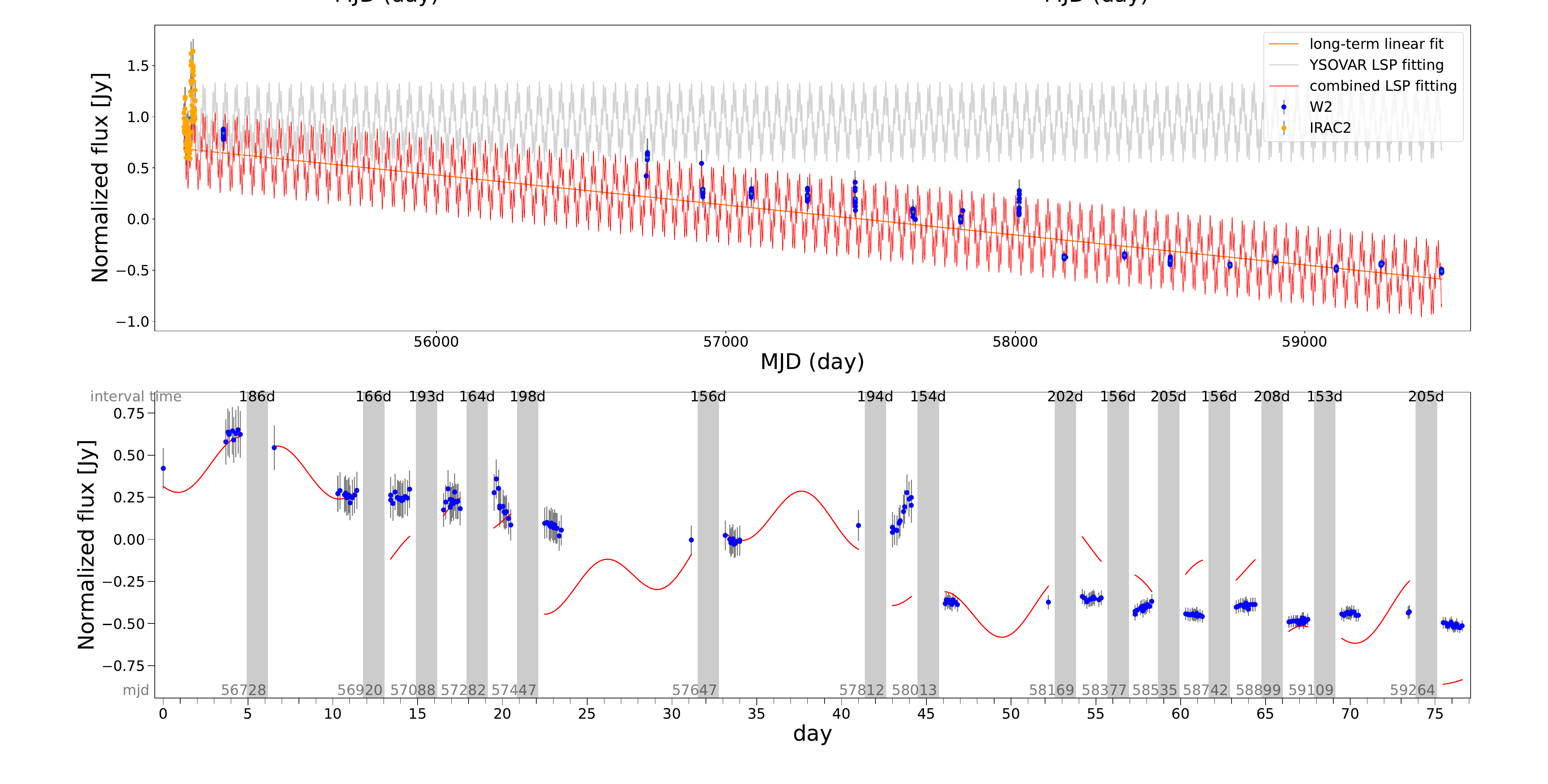} }
\caption{(Top) An example of lightcurve with YSOVAR IRAC2 (yellow), NEOWISE W2 data (blue) and the combined fitting of long-term with short-term (red line), corresponding to the same source as shown in Figure \ref{fig:veryshort1}.
(Bottom) NEOWISE very short-term lightcurve, where the data points with an interval of approximately 6 months are removed. The gray boxes indicate the interval time, marked at the top of each box, with '186d' indicating 186 days. The red line represents the combined fitting, which is the same as indicated in the top panel.
\label{fig:veryshort2}}
\end{figure*}

For instance, Figure \ref{fig:veryshort2} compares the observed and expected variability for the YSO V418 Ori. In the NEOWISE 7.5-year data, this source is classified as \textit{Linear+Irregular}, whereas in the YSOVAR 40-day timescale data, it is categorized as \textit{Curved+Periodic}. The bottom panel of the figure illustrates that the very short-term variability aligns well with the combined fitting trend. However, certain epochs exhibit discrepancies, which could arise from stochastic variability processes occurring at very short timescales, such as bursts resulting from unstable accretion \citep[see e.g.,][]{2017Cody}. Similar stochastic behavior can also be observed during the YSOVAR observations (refer to Figure \ref{fig:veryshort1}).

To assess the level of YSO variability across different timescales, we examined the variability amplitude ($\Delta$) at IRAC2 ($4.5\mu$m) and W2 ($4.6\mu$m) for the very short-term, short-term, and long-term periods. In the case of the very short-term analysis, we calculated the flux standard deviation (SD) within a specific epoch lasting 1-2 days. In this study, we chose one epoch, as an illustrative example, which occurred at MJD $\sim$ 57810, since data for all 703 sources that were cross-matched between the NEOWISE and YSOVAR catalogs (refer to Section \ref{subsec:4.1}) were available during this epoch. We conducted similar analyses for other very short-term epochs, and all yielded comparable results to those observed in this epoch. The short-term SD was determined based on the 40-day YSOVAR observations, while the long-term SD was derived from the 7.5-year NEOWISE data. To ensure the same sensitivity to variable identification, we adopt the mean flux uncertainty calculated from the NEOWISE data for all three timescales.

\begin{figure*}[ht!]
\centering
{\includegraphics[trim={3cm 1cm 4cm 3cm},clip,width=1\textwidth]
{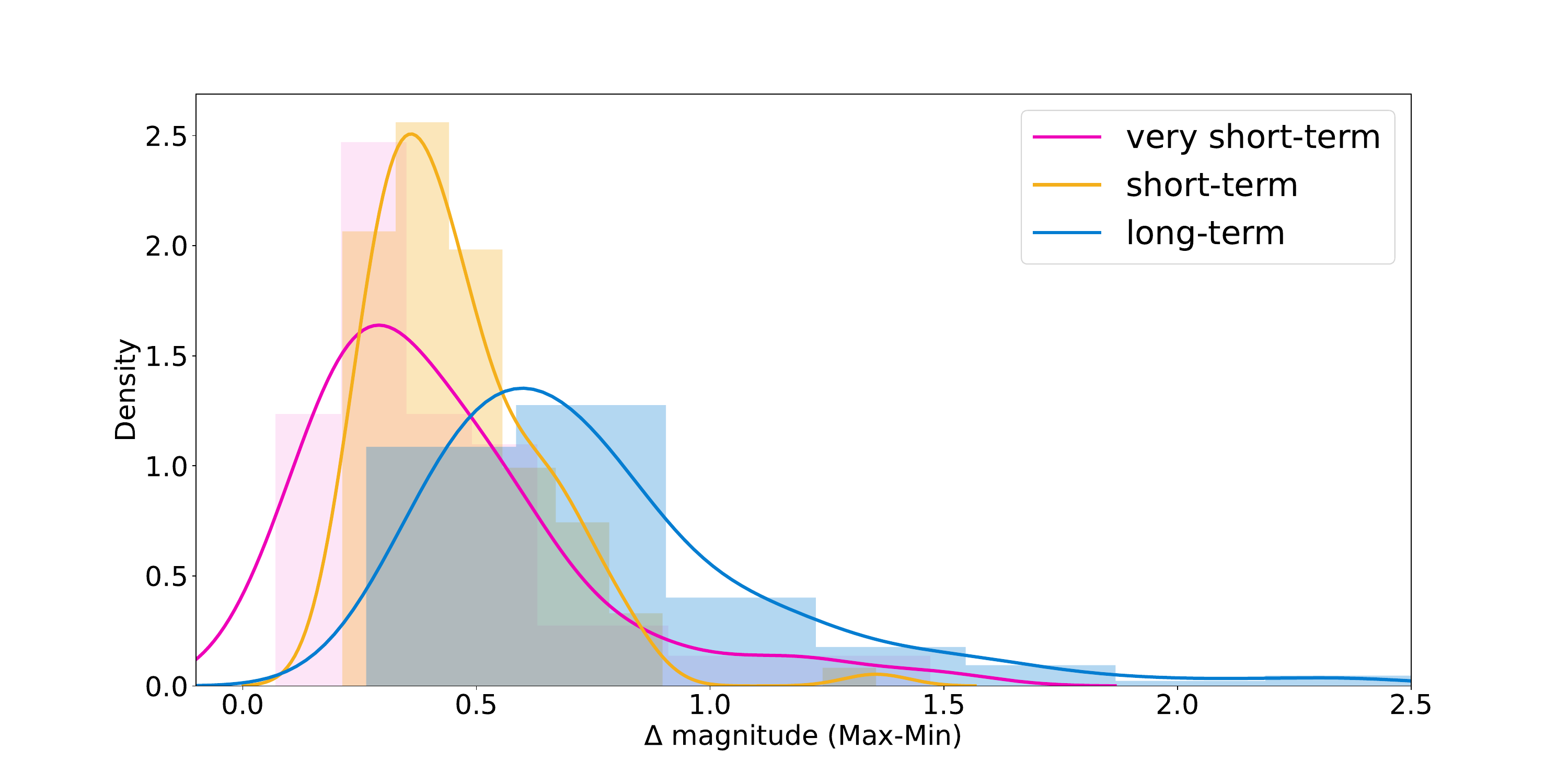} }
\caption{Histogram of $\Delta$ magnitude (Max - Min) at different timescales with Kernel Density Estimation (KDE) fitting. Each color corresponds to a specific timescale, highlighting the observed variability within each period. The histogram includes data for 264 long-term variables (blue), 106 short-term variables (yellow), and 52 very short-term variables (pink). The KDE fitting displays the peak value for each timescale, which are 0.3, 0.4, and 0.6, respectively. 
This indicates that the magnitude difference ($\Delta$ magnitude) tends to increase with longer timescales.
\label{fig:veryshort_hist}}
\end{figure*}

Figure \ref{fig:veryshort_hist} compares distributions of $\Delta$magnitude of variables in the three different timescales. Applying the same mean flux uncertainty of NEOWISE ($\sigma (W2)$) to both YSOVAR and NEOWISE data of the 703 cross-matched YSOs, variables were identified with the criteria of $SD/\sigma (W2) > 3$ in three different timescales: 52 very short-term variables from the NEOWISE measurements within the epoch of MJD $\sim$ 57810, 106 short-term variables from the YSOVAR data, and 264 long-term variables from the epoch-averaged NEOWISE data. 

A pair-wise Kolmogorov-Smirnov (KS) test, utilizing the scipy.stats.kstest function from the Python scipy package
\citep{2020SciPy-NMeth}
was performed between the very-short/short and short/long results to determine if the samples were drawn from a single distribution. In both cases this null hypothesis was definitely rejected ($p << 0.05$), arguing that the three observed distributions are independent.
The histogram illustrates an increase in the value of $\Delta$ as the observational baseline extends, aligning with the findings of the previous section and previous YSO variability studies \citep[e.g.,][]{2012Scholz,2019Contreras}. Furthermore, a modest increase in $\Delta$ between very short-term and short-term variability exists. 
The change, however, is very small with respect to the breadths of each distribution, suggesting that mechanisms driving 1-2 day variability are prevalent among YSOs.

\subsection{Interesting Individual sources} \label{subsec:5.3}

This section focuses on specific sources that exhibit intriguing lightcurve behavior across the various timescales analyzed in this study.

\begin{figure*}
\centering
{\includegraphics[trim={5cm 3cm 5cm 1cm},clip,width=1\textwidth]
{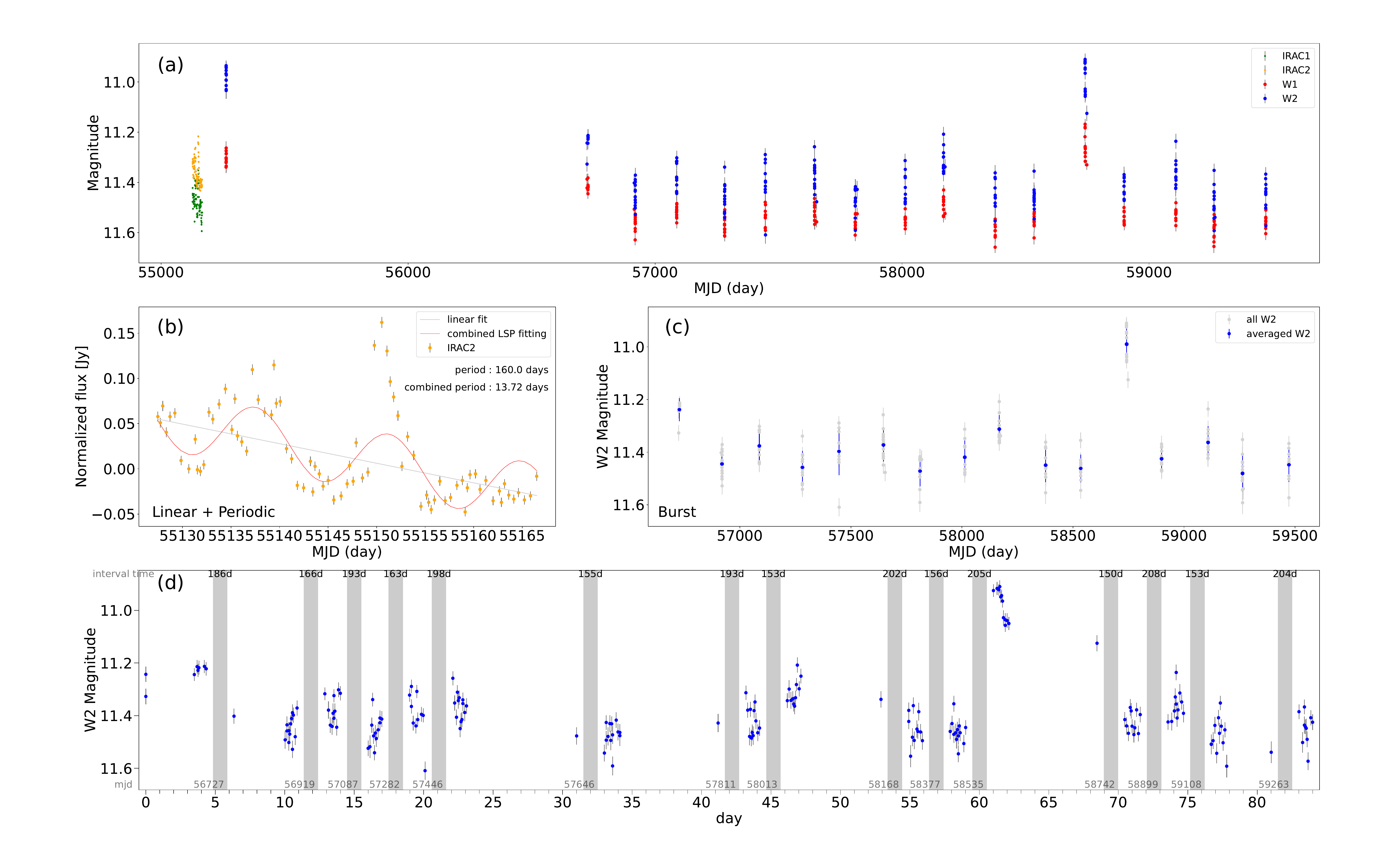} }
\caption{V935 Ori lightcurves in various timescales. It shows \textit{Linear+Periodic} with little burst in 40-day analysis (IRAC2, orange data) and \textit{Burst} in the long-term (W2, blue data).
\label{fig:1206}}
\end{figure*}

One such source is V935 Ori (Figure \ref{fig:1206}). Across the 40-day YSOVAR lightcurve it demonstrates a \textit{Linear + Periodic} trend with a combined period of 13.72 days. A burst-like event is observed around MJD $\sim$ 55150 and lasting for a few days, although it is not classified as a \textit{Burst} source due to the relatively small amplitude of the burst. In the NEOWISE long-term lightcurve, a more pronounced burst is evident, leading to its classification as a \textit{Burst} source. Additionally, the AllWISE data at approximately MJD 55200 also captures a burst event. These burst events are likely associated with unstable accretion resulting from the interaction between the stellar magnetosphere and the inner edge of the disk \citep{2014Stauffer, 2017Cody}.

\begin{figure*}[h]
\centering
{\includegraphics[trim={5cm 3cm 5cm 1cm},clip,width=1\textwidth]
{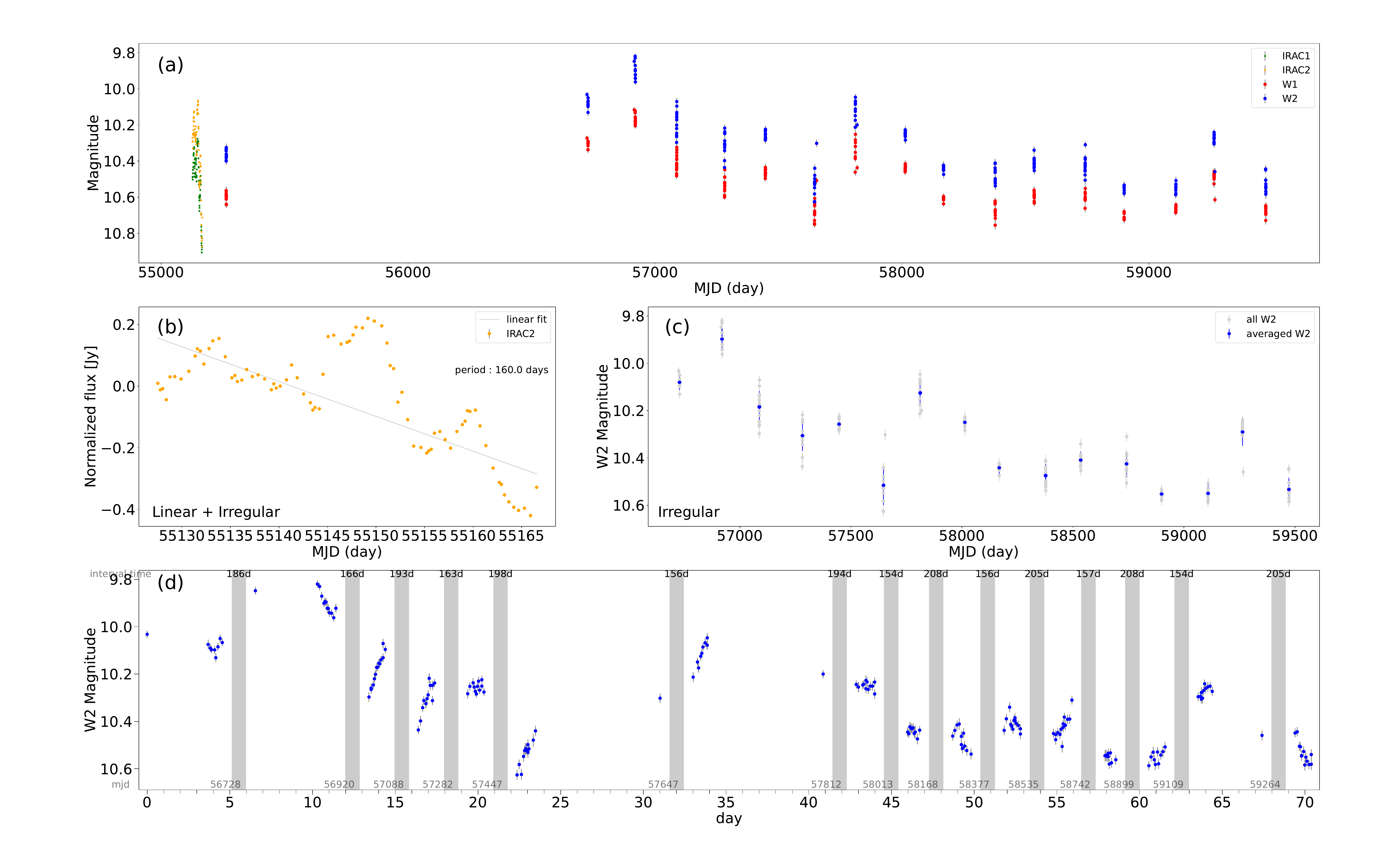} }
\caption{V2528 Ori lightcurves in various timescales. It shows \textit{Linear+Irregular} type in 40-day analysis (IRAC2, orange data) and \textit{Irregu lar} type in the long-term (W2, blue data).
\label{fig:2653}}
\end{figure*}

Another notable source, V2528 Ori (Figure \ref{fig:2653}), exhibits an \textit{Irregular} type in the long-term analysis, indicating irregular fluctuations over an extended period. However, in the 40-day analysis, it shows a \textit{Linear+Irregular} type, suggesting that the irregular fluctuations occur on top of a linear trend. The very short-term variability appears similar to the irregular fluctuations observed in the YSOVAR data. The complexity of the lightcurve in V2528 Ori is likely attributed to a combination of physical mechanisms driving its variability. The short-term irregularity, which appears nearly periodic on timescales exceeding a few days based on visual inspection, may be influenced by the presence of (hot) stellar spots. On the other hand, the dominant linear trend observed in the long-term analysis, as evidenced by the long period of approximately 160 days derived from the YSOVAR data, is most likely driven by structural changes occurring in the accretion disk of the YSO.

\begin{figure*}[ht!]
\centering
{\includegraphics[trim={5cm 3cm 5cm 1cm},clip,width=1\textwidth]
{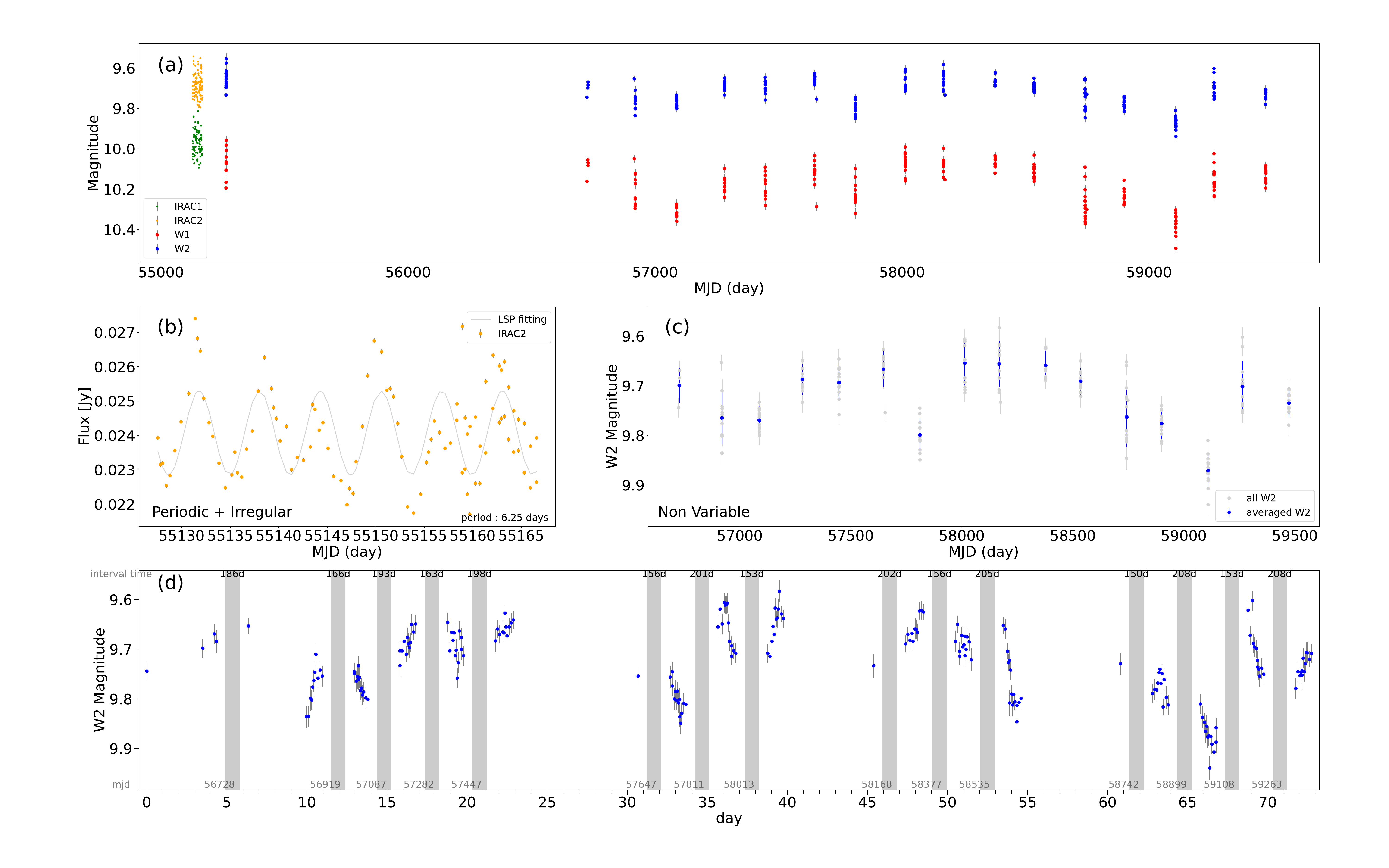} }
\caption{V2013 Ori lightcurves in various timescales. It shows \textit{Periodic+Irregular} type with a period of 6.25 days in 40-day analysis (IRAC2, orange data), but it is non-variable in the long-term (W2, blue data).
\label{fig:1597}}
\end{figure*}

A last example is V2012 Ori (JW 61) (Figure \ref{fig:1597}). In the long-term analysis, this source does not exhibit any variability. However, in the YSOVAR 40-day analysis, it is classified as \textit{Periodic}, displaying a periodic behavior with a period of 6.25 days. The very short-term data further confirms the persistence of this periodicity throughout the NEOWISE coverage of the source. In this case, the short-term periodic variability is most likely attributed to magnetic spots on the stellar photosphere. These spots, possibly associated with magnetic activity, give rise to the observed periodic variations in the lightcurve. However, it is worth noting that V2012 Ori is a binary system \citep{reipurth07}, which could potentially contribute to its variability.


By examining YSO variability across these different timescales, we can gain a more comprehensive understanding of the physical processes driving the observed variability. This multi-timescale approach helps us to distinguish between different types of variability, identify correlations between short-term and long-term behaviors, and unravel the complex interplay of various mechanisms influencing YSOs' brightness variations.

\section{SUMMARY} \label{sec:6}

Variability in YSOs is a common phenomenon that can be caused by various physical mechanisms affecting the stellar photosphere, accretion disk and envelope in these systems. To gain a better understanding of these mechanisms it is important to examine YSO variability on different timescales. In particular, mid-IR data sets are powerful for evaluating protostellar accretion as well as other physical mechanisms driving YSO variability.

In this study, we utilize mid-IR data arising from the NEOWISE and YSOVAR surveys, which cover the wavelength range of 3 to 5 $\mu$m over 7.5 years and 40-day timescales respectively. From analysis of these data we found.
 
\begin{enumerate}

  \item The majority of YSOs exhibit variability on the 40-day timescales provided by YSOVAR (77.6\%). This variability is primarily driven by YSOs classified as secular variables, i.e objects that display variability over timescales larger than 20 days. On the other hand, less than half of the YSOs (43.0\%) display variability on the 7.5 year timescales from NEOWISE, primarily attributed to objects classified as stochastic variables (Section \ref{sec:3}).
  
  \item  We found 703 YSOs in Orion which contain data sets from both surveys.  Among these YSOs, a higher percentage (93.0\%) exhibited variability in the short-term, while a lower percentage (41.8\%) showed variability in the long-term (Section \ref{subsec:4.1}). Additionally, the amplitude of variability in the short-term was smaller than in the long-term, suggesting that the stellar photosphere-inner disk interface causes the short-term variability, while accretion or geometrical variations in the disk result in the long-term variability (Section \ref{subsec:4.2}).
 
  \item We studied the possible effect that the lower sensitivity of NEOWISE data, compared with YSOVAR observations, could have on our results. When we degraded the YSOVAR data using the NEOWISE magnitude error, the fraction of variable sources significantly decreased from 81.9\% to 15.1\%. This suggests that the sensitivity of the telescope is crucial in detecting YSO variability (Section \ref{subsec:4.3}). 

  \item Finally, we studied the YSO variability in a very short-term timescale (1-2 days) provided by the single-epoch NEOWISE observations. We find that the variability amplitude ($\Delta$) increases as the observational baseline extends (Section \ref{subsec:5.2}), in agreement with previous studies of the variability of YSOs.

\end{enumerate}

\appendix
\section{NEOWISE Outlier data cut} \label{app:1}

In our analysis of the 7.5-year NEOWISE data, we have adopted a different approach compared to the 6.5-year NEOWISE data analysis conducted in \citetalias{park21}. In \citetalias{park21}, they utilized only 70\% of the exposures in each epoch to mitigate potential errors in the photometry. However, considering that most YSOs exhibit variability on short timescales \citep{morales11}, removing 30\% of the NEOWISE data as done in \citetalias{park21} could potentially hinder our ability to detect very short-term variability within a single NEOWISE epoch. 

To address this concern, we have employed an alternative method to select high-quality exposures based on the photometry parameters provided in the NEOWISE database. Specifically, we have considered parameters such as the signal-to-noise ratio (SNR) of each exposure and the distance to the coordinates of the YSO (see Section \ref{subsec:2.1}). By applying these selection criteria, we ensure that the data used for analyzing very short-term variability are of sufficient quality.

\begin{figure}[h]
\centering
{\includegraphics[trim={0cm 0cm 0cm 0cm},clip,width=1\textwidth]
{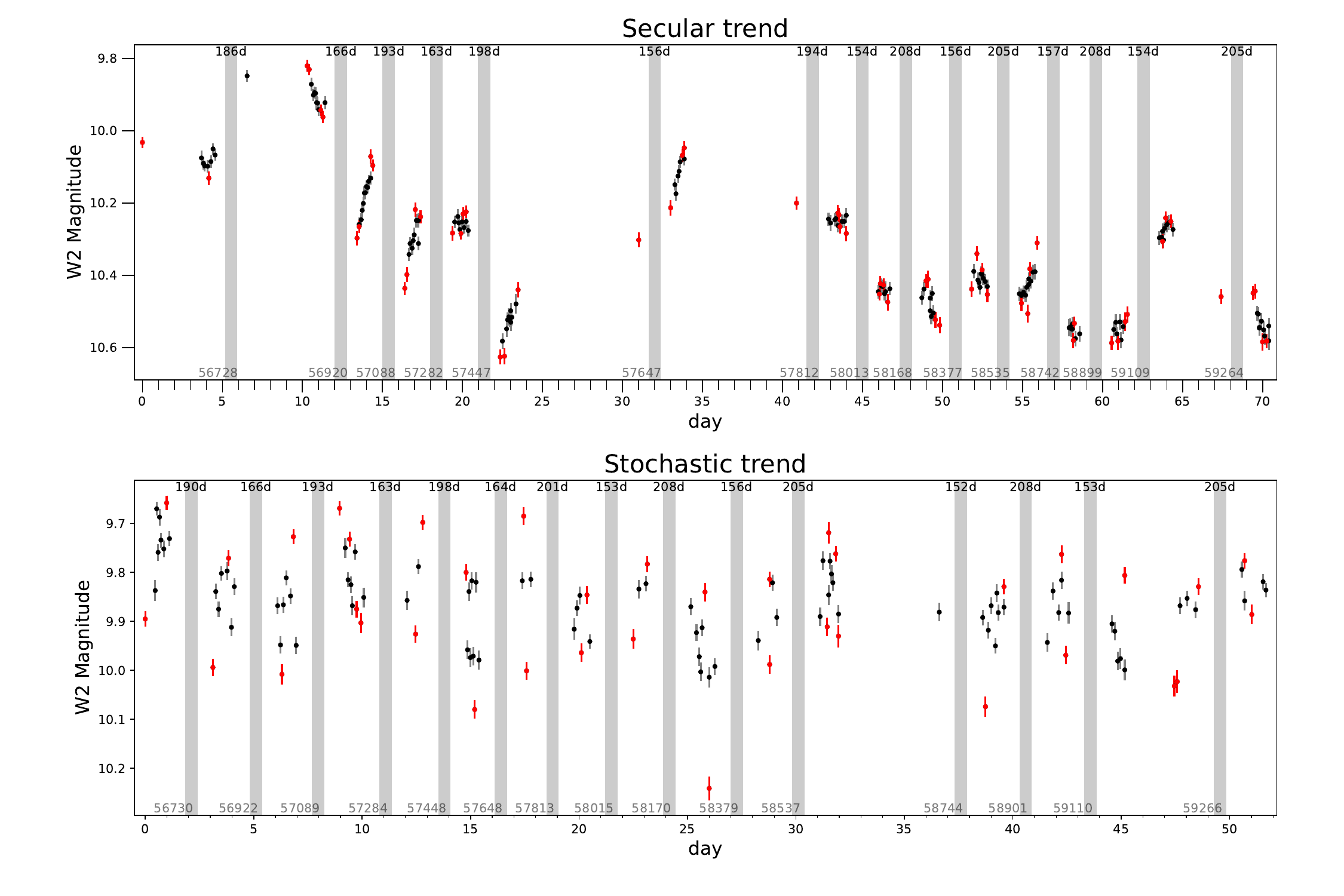} }
\caption{Examples of very short-term lightcurve with marked 30\% Outlier (red). (Top) Lightcurve of a source showing secular trends in the very short-term variable. Most of the variables have this trend. (Bottom) Lightcurve of a source showing stochastic trends in the very short-term variable.
\label{fig:outlier}}
\end{figure}

Figure \ref{fig:outlier} illustrates two lightcurves, comparing the results obtained by removing 30\% of the NEOWISE data versus our approach of selecting high-quality exposures. We have observed that in most sources, removing the 30\% of data would result in the loss of data points that exhibit genuine variability at very short timescales.

Therefore, instead of discarding a fixed percentage of data, we opted for a data selection strategy based on the photometry parameters in the NEOWISE database. This approach allows us to retain valuable information regarding very short-term variability while ensuring the reliability and quality of the data used in our analysis.

\section*{Acknowledgement}

This work was supported by the National Research Foundation of Korea (NRF) grant funded by the Korea government (MSIT) (grant number 2021R1A2C1011718). G.J.H.\ is supported by general grants 12173003 awarded by the National Natural Science Foundation of China.
D.J.\ is supported by NRC Canada and by an NSERC Discovery Grant.  

This publication makes use of data products from NEOWISE, which is a project of the Jet Propulsion Laboratory/California Institute of Technology, funded by the Planetary Science Division of the National Aeronautics and Space Administration.

\bibliography{reference}{}
\bibliographystyle{aasjournal}

\end{document}